\documentclass[preprint,amsfonts,aps]{revtex4}
\usepackage{graphicx}
\usepackage{amssymb}

\begin{document}
\title{The arrow of time: from universe time-asymmetry to local irreversible
processes}

\author{Mat\'\i as Aiello}
\email{aiello@iafe.uba.ar}
\thanks{ANPCyT Fellow}
\affiliation{Instituto de  Astronom\'\i a y F\'\i sica del
Espacio, Casilla de Correo 67, Sucursal 28, 1428 Buenos Aires,
Argentina} \affiliation{Departamento de F\'\i sica, Facultad de
Ciencias Exactas y Naturales, Universidad de Buenos Aires, Ciudad
Universitaria, Pabell\'on I, 1428 Buenos Aires, Argentina}

\author{Mario Castagnino}
\email{mariocastagnino@citynet.net.ar} \thanks{CONICET}
\affiliation{Instituto de F\'{i}sica de Rosario}
\affiliation{Instituto de Astronom\'{i}a y F\'{i}sica del Espacio,
Casilla de Correos 67, Sucursal 28, 1428, Buenos Aires, Argentina}

\author{Olimpia Lombardi}
\email{olimpiafilo@arnet.com.ar}
\thanks{CONICET-IECT}
\affiliation{CONICET- Universidad de Buenos Aires\\
C. Larralde 3440, 6${{}^{o}}$D, 1430, Buenos Aires, Argentina. }

\begin{abstract}
In several previous papers we have argued for a global and non-entropic
approach to the problem of the arrow of time, according to which the
''arrow'' is only a metaphorical way of expressing the geometrical
time-asymmetry of the universe. We have also shown that, under definite
conditions, this global time-asymmetry can be transferred to local contexts
as an energy flow that points to the same temporal direction all over the
spacetime. The aim of this paper is to complete the global and non-entropic
program by showing that our approach is able to account for irreversible
local phenomena, which have been traditionally considered as the physical
origin of the arrow of time.
\end{abstract}

\maketitle

\tableofcontents
\newpage
\thispagestyle{empty} \cleardoublepage

\section{Introduction}

In several previous papers$^{(1,2,3,4,5,6)}$ we have argued for a global and
non-entropic approach to the problem of the arrow of time, according to
which the ''arrow'' is only a metaphorical way for expressing the
geometrical time-asymmetry of the universe. We have also shown that, under
definite conditions, this global time-asymmetry can be transferred to local
contexts as an energy flow that points to the same temporal direction all
over the spacetime. However, many relevant irreversible local phenomena were
still unexplained by our approach. The account of them is necessary to reach
a full answer to the problem of the arrow of time, since they have been
traditionally considered as the physical origin of such an arrow. The aim of
this paper is to complete the global and non-entropic program by showing
that our approach is able to account for those local irreversible phenomena.

For this purpose, the paper is organized as follows. In Section II we
introduce the precise definition of the basic concepts involved in the
discussion: time-reversal invariance, irreversibility and arrow of time. In
Section III we summarize our global and non-entropic approach to the problem
of the arrow of time, according to which the arrow is given by the
time-asymmetry of spacetime. In this section we also explain how the global
arrow is transferred to local contexts as an energy flow defined all over
the spacetime and which, as a consequence, represents a relevant physical
magnitude in local theories. Section IV is devoted to show how the energy
flow breaks the time-symmetry of the pair of solutions, one the temporal
mirror image to the other, resulting from different time-reversal invariant
fundamental laws. In particular, we consider quantum mechanics, quantum
field theory, and the case of Feynman graphs and quantum measurements. In
section V, irreversibility at the phenomenological level is discussed: we
show that, when phenomenological theories are analyzed in fundamental terms,
a second irreversible solution evolving towards the past can always be
identified; the energy flow is what breaks the just discovered time-symmetry
of the pair. Finally, in Section VI we draw our conclusions.

\section{Basic concepts}

It is surprising that, after so many years of debates about irreversibility
and time's arrow, the meanings of the terms involved in the discussion are
not yet completely clear: the main obstacle to agreement is conceptual
confusion. For this reason, we begin with disentangling the basic concepts
of the problem.

\subsection{Time-reversal invariance}

Even a formal concept as time-reversal invariance is still object of
controversies (see Albert's recent book$^{(7)}$, and Earman's criticisms$%
^{(8)}$). We define it as follows:

\begin{quote}
{\bf Definition 1: }A dynamical equation is {\it time-reversal invariant} if
it is invariant under the application of the time-reversal operator ${\cal T}
$, which performs the transformation $t\rightarrow -t$ and reverses all the
dynamical variables whose definitions in function of $t$ are non-invariant
under the transformation $t\rightarrow -t$.
\end{quote}

On the basis of this definition, we can verify by direct calculation that
the dynamical equations of fundamental physics are time-reversal invariant.
Let us see some examples:

\begin{itemize}
\item  {\bf Classical mechanics:} In ordinary classical mechanics, the basic
magnitudes (position ${\bf x}$, velocity ${\bf v}$ and acceleration ${\bf a}$%
) change as
\begin{equation}
{\cal T}{\bf x}={\bf x},\text{ \quad }{\cal T}{\bf v}=-{\bf v},\quad {\cal T}%
{\bf a}={\bf a}  \label{1}
\end{equation}

In general, mass $m$ and force ${\bf F}$ are conserved magnitudes, that is,
they are not functions of $t$; then,
\begin{equation}
{\cal T}m=m,\quad {\cal T\,}{\bf F}={\bf F}  \label{2}
\end{equation}

Since ${\bf F}=-{\bf \nabla }V({\bf x})$, where $V$ is a potential, the
energy $H=\frac{1}{2}m{\bf v}^{2}+V({\bf x})$ is invariant under the action
of ${\cal T}$:
\begin{equation}
{\cal T}H=H  \label{3}
\end{equation}
Analogously, in Hamiltonian classical mechanics, the position ${\bf q}%
=(x_{1},...,x_{n})$ and the momentum ${\bf p}=(p_{1},...,p_{n})$ change as
\begin{equation}
{\cal T}{\bf q}={\bf q},\text{ \quad }{\cal T}{\bf p}=-{\bf p}  \label{4}
\end{equation}

\item  {\bf Electromagnetism:} Here the charge $q$ is not function of $t$
since it is also a conserved magnitude; so,
\begin{equation}
{\cal T}q=q  \label{5}
\end{equation}
Then, the charge density $\rho $ and the current density ${\bf j=}\rho {\bf v%
}$ change as
\begin{equation}
\text{ }{\cal T}\rho =\rho ,\text{ \quad }{\cal T\,}{\bf j}=-{\bf j}
\label{6}
\end{equation}

Since the Lorentz force is defined as ${\bf F}=q{\bf E}+{\bf j}\times {\bf B}
$, from eqs. (\ref{2}), (\ref{5}) and (\ref{6}) the electric field ${\bf E}$
and the magnetic induction ${\bf B}$ change as
\begin{equation}
{\cal T\,}{\bf E}={\bf E,\quad }{\cal T\,}{\bf B}=-{\bf B}  \label{7}
\end{equation}

\item  {\bf Quantum mechanics: }In order to apply the time-reversal operator
to quantum mechanics, the configuration representation has to be used: $%
x\sim x,\quad $ $p\sim -i\hbar \frac{\partial }{\partial x}$. Since we want
to obtain ${\cal T}x=x$ and ${\cal T}p=-p$ as in the classical case, we
impose that the wave function change as ${\cal T}\phi (x)=\phi (x)^{*}$. But
this requirement makes the quantum time-reversal operator {\it antilinear }%
and {\it antiunitary }(by contrast with the linearity of ${\cal T}$ in
classical mechanics). In order to express this difference, the time-reversal
operator in quantum mechanics is denoted by ${\bf T}$:
\begin{equation}
{\bf T}\phi (x)=\phi (x)^{*}  \label{8}
\end{equation}
In fact, if $\omega $ are the eigenvalues of the Hamiltonian $H$, with the
linear ${\cal T}$ we would obtain ${\cal T}e^{-i\omega t}=e^{i\omega t}$
and, therefore, ${\cal T}\omega =-\omega $, which would lead to unacceptable
negative energies. On the contrary, with the antilinear ${\bf T}$,
\begin{equation}
{\bf T}e^{-i\omega t}=e^{-i\omega t},\quad {\bf T}H=H,\quad {\bf T}\omega
=\omega   \label{9}
\end{equation}

\item  {\bf Quantum field theory:} In this chapter of physics, the linear
and unitary operator ${\bf {P}}$, corresponding to space-inversion, and the
antilinear and antiunitary operator ${\bf {T}}$, corresponding to
time-reversal, apply to the quadri-momentum $P^{\mu }$ as (we will return on
this point in Section IV.B)
\begin{equation}
{\bf P}iP^{\nu }{\bf P}^{-1}=i{\cal P}_{\mu }^{\nu }P^{\mu },\quad {\bf T}%
iP^{\nu }{\bf T}^{-1}=i{\cal T}_{\mu }^{\nu }P^{\mu }  \label{10}
\end{equation}
where
\begin{equation}
{\cal P}_{\nu }^{\mu }=\left(
\begin{array}{rrrr}
1 & 0 & 0 & 0 \\
0 & -1 & 0 & 0 \\
0 & 0 & -1 & 0 \\
0 & 0 & 0 & -1
\end{array}
\right) ,\qquad {\cal T}_{\nu }^{\mu }=\left(
\begin{array}{rrrr}
-1 & 0 & 0 & 0 \\
0 & 1 & 0 & 0 \\
0 & 0 & 1 & 0 \\
0 & 0 & 0 & 1
\end{array}
\right)   \label{11}
\end{equation}
\end{itemize}

As a consequence of the definition of time-reversal invariance, given a
time-reversal invariant equation $L$, if $f(t)$ is a solution of $L$, then $%
{\cal T}f(t)$ is also a solution. In previous papers$^{(2,4)}$, we have
called these two mathematical solutions ''{\it time-symmetric twins}'': they
are twins because, without presupposing a privileged direction of time, they
are only conventionally different; they are time-symmetric because one is
the temporal mirror image of the other. The traditional example of
time-symmetric twins is given by electromagnetism, where dynamical equations
always have advanced and retarded solutions, respectively related with
incoming and outgoing states in scattering as described by Lax-Phillips
theory$^{(9)}$. The two twins are identical and cannot be distinguished at
this stage since, up to now, there is no further criterion than the
time-reversal invariant dynamical equation from which they arise.
Conventionally we can give a name to each solution: ''advanced'' and
''retarded'', ''incoming'' and ''outgoing'', etc. But these names are just
conventional labels and certainly do not establish a non-conventional
difference between both time-symmetric solutions.

In general, the dynamical equations of {\it fundamental} physics are
time-reversal invariant, e.g. the dynamical equation of classical mechanics,
the Maxwell equations of electromagnetism, the Schr\"{o}dinger equation of
quantum mechanics, the field equations of quantum field theory, the Einstein
field equations of general relativity. However, not all axioms of
fundamental theories are time-reversal invariant; this is the case of the
Postulate III of quantum field theory (see Section IV.B) and the measurement
postulate of quantum mechanics (see Section IV.C). On the other hand, many
non fundamental laws are non time-reversal invariant, as the
phenomenological second law of thermodynamics (see Section V). One of the
purposes of this paper is to explain these apparent ''anomalies{\it ''}.

\subsection{Irreversibility}

Although the concepts of reversibility and irreversibility have received
many definitions in the literature on the subject, from a very general
viewpoint a reversible evolution is usually conceived as a process that can
occur in the opposite temporal order according to the dynamical law that
rules it: the occurrence of the opposite process is not excluded by the law.
The typical irreversible processes studied by physics are decaying
processes, that is, time evolutions that tend to a final equilibrium state
from which the system cannot escape: the irreversibility of the process is
due precisely to the fact that the evolution leaving the equilibrium state
is not possible. For these cases, reversibility can be defined as:

\begin{quote}
{\bf Definition 2: }A solution $f(t)$ of a dynamical equation is {\it %
reversible} if it does not reach an equilibrium state (namely, if $\nexists
\lim_{t\rightarrow \infty }f(t))$ where the system remains forever.
\end{quote}

For instance, according to this definition, in classical mechanics a
solution of a dynamical equation is reversible if it corresponds to a closed
curve in phase space (even if these curves are closed through a point at
infinite); if not, it is irreversible.

\begin{figure}[h]
\begin{center}
\includegraphics[width=6cm,angle=0]{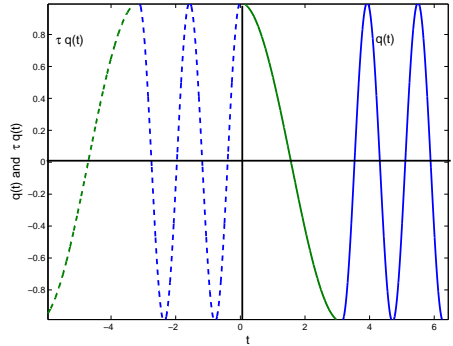}
\caption{Time-asymmetric solutions for $q(t)$} \label{osc}
\end{center}
\end{figure}

It is quite clear that time-reversal invariance and reversibility are
different concepts to the extent that they apply to different mathematical
entities: time-reversal invariance is a property of dynamical equations and,
{\it a fortiori}, of the set of its solutions; reversibility is a property
of a single solution of a dynamical equation. Furthermore, they are not even
correlated, since both properties can combine in the four possible cases
(see Castagnino, Lara and Lombardi$^{(2)}$). In fact, besides the usual
cases time-reversal invariance-reversibility and non time-reversal
invariance-irreversibility, the remaining two combinations are also possible:

\begin{itemize}
\item  {\bf Time-reversal invariance and irreversibility. }Let us consider
the pendulum with Hamiltonian
\begin{equation}
H=\frac{1}{2m}\,p_{\theta }^{2}-\frac{k^{2}}{2}\,\cos \theta   \label{12}
\end{equation}
The dynamical equations are time-reversal invariant since ${\cal T\,}\theta
=\theta $:
\begin{equation}
\dot{\theta }=\frac{\partial H}{\partial p_{\theta }}=\frac{%
p_{\theta }}{m}\qquad \qquad \dot{p_{\theta }}=-\frac{%
\partial H}{\partial \theta }=-\frac{k^{2}}{2}\sin \theta   \label{13}
\end{equation}

Therefore, the set of trajectories in phase space is symmetric with respect
to the $\theta $-axis. However, not all the solutions are reversible. In
fact, when $H=\frac{k^{2}}{2}$, the solution is irreversible since it tends
to $\theta =\pi $, $p_{\theta }=0$ ($\theta =-\pi $, $p_{\theta }=0$) when $%
t\rightarrow \infty $ ($t\longrightarrow -\infty $) (see Tabor$^{(10)}$): it
corresponds to the pendulum reaching its unstable equilibrium state for $%
t\rightarrow \infty $ ($t\longrightarrow -\infty $). For $H<\frac{k^{2}}{2}$
(oscillating pendulum) and $H>\frac{k^{2}}{2}$ (rotating pendulum), the
evolutions are reversible.

\item  {\bf Non time-reversal invariance and reversibility. }Let us now
consider the modified oscillator with Hamiltonian
\begin{equation}
H=\frac{1}{2m}\,p^{2}+\frac{1}{2}\,K(p)^{2}\,q^{2}  \label{14}
\end{equation}
where $K(p)=K_{+}$ {\tiny \ }when {\tiny \ }$p\geq 0$, $K(p)=K_{-}${\tiny \ }%
when {\tiny \ }$p<0$, and $K_{+}$ and $K_{-}$ are constants. This means that
${\cal T\,}K_{+}=K_{-}$. As a consequence, if $K_{+}\neq K_{-}$, the
dynamical equations are non time-reversal invariant since, for $p\geq 0$,
\begin{equation}
\dot{p}=-K_{+}^{2}q\quad \,{\cal T}\dot{p}=-%
{\cal T\,}K_{+}^{2}\,{\cal T\,}q=-K_{-}^{2}\,q\neq -K_{+}^{2}q  \label{15}
\end{equation}
and for $p<0$,
\begin{equation}
\dot{p}=-K_{-}^{2}q\qquad {\cal T}\dot{p}=-%
{\cal T\,}K_{-}^{2}\,{\cal T\,}q=-K_{+}^{2}\,q\neq -K_{-}^{2}q  \label{16}
\end{equation}

Nevertheless, the solutions $q(t)$ and $p(t)$ are, for $p\geq 0$,
\begin{equation}
q(t)=C_{1}\cos (\omega _{+}t+\alpha _{+n})\quad p(t)=C_{1}m\omega _{+}\sin
(\omega _{+}t+\alpha _{+n})  \label{17}
\end{equation}
and for $p<0$,
\begin{equation}
q(t)=C_{2}\cos (\omega _{-}t+\alpha _{-n})\quad p(t)=C_{2}m\omega _{-}\sin
(\omega _{-}t+\alpha _{-n})  \label{18}
\end{equation}

where $\omega _{\pm }=\frac{K_{\pm }^{2}}{m}$ and the constants
$\alpha _{\pm n}$ change from one cycle $n$ to the next cycle
$n+1$ in such a way that the solutions turn out to be continuous.
In Fig. \ref{osc} we display the time-asymmetric solutions $q(t)$
for this example. It is clear that these solutions have no limit
for $t\rightarrow \pm \infty $: each trajectory is reversible
since it is a closed curve in phase space.
\end{itemize}

Once both concepts are elucidated in this way, {\it the problem of
irreversibility} can be clearly stated: {\it how to explain irreversible
evolutions in terms of time-reversal invariant laws}. When explained in
these terms, it turns out to be clear that there is no conceptual puzzle in
the problem of irreversibility: nothing prevents a time-reversal invariant
equation from having irreversible solutions. Nevertheless, the solution of
the problem of irreversibility does not provide yet an adequate distinction
between the two directions of time. In fact, if an irreversible evolution is
a solution of a time-reversal invariant law, there will always exist its
time-symmetric twin, that is, another irreversible solution that is its
temporal mirror image. For instance, if there is an irreversible solution
leading to equilibrium towards the future, necessarily there exists another
irreversible solution leading to equilibrium towards the past, and there is
no non-conventional criterion for selecting one of the temporally opposite
evolutions as the physically relevant.

In general, a privileged direction of time is presupposed when irreversible
processes are studied. In fact, when we talk about entropy increasing
processes, we suppose an entropy increase {\it towards the future}; or when
we consider a process going from non-equilibrium to equilibrium, we
implicitly locate equilibrium {\it in the future}. In general, any evolution
that tends to an attractor is conceived as approaching it towards the
future. This means that the distinction between past and future is usually%
{\it \ taken for granted}, and this fact usually hides the existence of the
second irreversible twin of the pair. However, when the time-reversal
invariant theory is developed without projecting our time-asymmetric
intuitions, the pair of time-symmetric twins becomes manifest.

\subsection{Arrow of time}

The problem of the arrow of time owes its origin to the intuitive asymmetry
between past and future. We experience the time order of the world as
''directed'': if two events are not simultaneous, one of them is earlier
than the other. Moreover, we view our access to past and future quite
differently: we remember past events and predict future events. On the other
hand, we live in a world full of processes that never occur in the opposite
direction: coffee and milk always mix together, we always get older, and
regrettably we never see the reversed processes. Therefore, if we conceive
the problem of the arrow of time as the question ''Does the arrow of time
exist?'', we can legitimately solve it on the basis of our best grounded
experiences: there is a non merely conventional difference between the two
directions of time, and the privileged direction, that we call ''future'',
is the direction of those well known processes.

However, this is not the problem of the arrow of time as conceived in the
foundations of physics since the birth of thermodynamics. In this context,
the difficulty consists in finding a {\it physical correlate} of the
experienced difference between the two temporal directions. If such a
temporal asymmetry did not exist, there would be no need to ask physics for
its explanation. It is precisely due to the ''directedness'' of our
experience of time that we want to find this feature accounted for by
physical theories. But, then, we cannot project our time-asymmetric
experiences and observations into the solution of the problem without
begging the question. In this paper, we will address the problem of the
arrow of time within the limits of physics: we will not discuss our
experiences about time and processes. Our question will be: {\it Do physical
theories pick out a preferred direction of time}?

The main difficulty to be encountered in answering this question relies on
our anthropocentric perspective: the difference between past and future is
so deeply rooted in our language and our thoughts that it is very difficult
to shake off these temporally asymmetric assumptions. In fact, traditional
discussions around the problem of the arrow of time in physics are usually
subsumed under the label ''the problem of the direction of time'', as if we
could find an exclusively physical criterion for singling out the privileged
direction of time, identified with what we call ''future''. But there is
nothing in the dynamical laws of physics that distinguishes, in a
non-arbitrary way, between past and future as we conceive them in our
ordinary language and our everyday life. It might be objected that physics
implicitly assumes this distinction with the use of temporally asymmetric
expressions, like ''future light cone'', ''initial conditions'',
''increasing time'', and so on. However this is not the case, and the reason
relies on the distinction between ''conventional'' and ''substantial''.

\begin{quote}
{\bf Definition 3: }Two objects are {\it formally identical} when there is a
permutation that interchanges the objects but does not change the properties
of the system to which they belong.
\end{quote}

In physics it is usual to work with formally identical objects: the two
semicones of a light cone, the two spin directions, etc.

\begin{quote}
{\bf Definition 4:} We will say that we establish a {\it conventional}
difference between two objects when we call two formally identical objects
with two different names.
\end{quote}

This is the case when we assign different signs to the two spin directions,
or different names to the two light semicones of a light cone, etc.

\begin{quote}
{\bf Definition 5:} We will say that the difference between two objects is
{\it substantial} when we assign different names to two objects that are not
formally identical. In this case, although the particular names we choose
are conventional, the difference is substantial (see Penrose$^{(11)}$, Sachs$%
^{(12)}$).
\end{quote}

For instance, the difference between the two poles of the theoretical model
of a magnet is conventional since both poles are formally identical; on the
contrary, the difference between the two poles of the Earth is substantial
because in the North Pole there is an ocean and in the South Pole there is a
continent (and the difference between ocean and continent remains
substantial even if we conventionally change the names of the poles). In
mathematics, given a segment $[\overline{A,B}]$, if we call its end points ''%
$A"$ and ''$B"$ (or ''$B"$ and ''$A"$, since the names are always
conventional), we are establishing a conventional difference between them
because the points $A$ and $B$ are formally identical; on the contrary, if
we call the end points of an arrow $\overrightarrow{[A,B]}$ ''$A$'' and ''$B$%
'', we are expressing a substantial difference between both points since the
tail $A$ is not formally identical to the head $B$.

Once this point is accepted, it turns out to be clear that, given the
time-reversal invariance of the fundamental laws, physics uses the labels
''past'' and ''future'' in a conventional way. Therefore, the problem cannot
be posed in terms of identifying the privileged direction of time named
''future'', as we conceive it in our ordinary language: the problem of the
arrow of time in physics becomes the problem of finding a {\it substantial
difference} {\it between the two temporal directions grounded only on
physical theories}. But if this is our central question, we cannot project
our experiences about past and future for solving it. When we want to
address the problem of the arrow of time from a perspective purged of our
temporal intuitions, we must avoid the conclusions derived from subtly
presupposing temporally asymmetric notions. As Huw Price$^{(13)}$ claims, it
is necessary to stand at a point outside of time, and thence to regard
reality in atemporal terms: this is his ''{\it view from nowhen}''. This
atemporal standpoint prevents us from using temporally asymmetric
expressions in a non-conventional way: the assumption about the difference
between past and future is not yet legitimate in the context of the problem
of the arrow of time.

But then, what does ''the arrow of time'' mean when we accept this
constraint? Of course, the traditional expression coined by Eddington has
only a metaphorical sense: its meaning must be understood by analogy. We
recognize the difference between the head and the tail of an arrow on the
basis of its intrinsic properties; therefore, we can substantially
distinguish between both directions, head-to-tail and tail-to-head, {\it %
independently of our particular perspective and our pretheoretical
intuitions.} Analogously, we will conceive {\it the problem of the arrow of
time} in terms of {\it the possibility of establishing a substantial
distinction between the two directions of time on the basis of exclusively
physical arguments}.

\section{The problem of the arrow of time: a global and non-entropic approach
}

On the basis of the distinction between conventional and substantial
differences, and of the need of an atemporal standpoint, we have proposed
and developed a global and non-entropic approach to the problem of the arrow
of time in several previous papers. Here we will only summarize the main
points of our argument.

\subsection{Why global and non-entropic}

Let us begin with explaining in what sense our approach moves away from the
traditional local and entropic way of addressing the problem.\bigskip

{\bf 1. Why global?\smallskip }

The traditional local approach owes its origin to the attempts to reduce
thermodynamics to statistical mechanics: in this context, the usual answer
to the problem of the arrow of time consists in defining the future as the
direction of time in which entropy increases. However, already in 1912 Paul
and Tatiana Ehrenfest$^{(14)}$ noted that, when entropy is defined in
statistical terms on the underlying classical dynamics, if the entropy of a
closed system increases towards the future, such increase is matched by a
similar one towards the past: if we trace the evolution of a non-equilibrium
system back into the past, we obtain states closer to equilibrium. This old
discussion can be generalized to the case of any kind of evolution arising
from local time-reversal invariant laws. In fact, as we have seen in the
previous section, any time-reversal invariant equation gives rise to a pair
of time-symmetric twins $f(t)$ and ${\cal T}f(t)$, which are only
conventionally different to each other.

Of course, the existence of time-symmetric twins is a result of a formal
property of the equation. When it represents a local dynamical law, the
solutions are conceived as representing two possible evolutions relative to
that law, because local physics assumes a one-to-one mapping between
possible evolutions and solutions of the dynamical law. But since both
solutions are only conventionally different, they do not supply a
substantial distinction between the two directions of time. In the face of
this problem, one might be tempted to solve it by simply stating that both
solutions describe the same process from temporally reversed viewpoints.
However, the fact that a single process be described by $e(t)$ and $e(-t)$
means that ''$t$'' and ''$-t$'' are only two different names for the same
temporal point. Therefore, the time represented by $[0,\infty )$ and the
time represented by $(-\infty ,0]$ would be not conventionally different,
but strictly identical. Then, time itself would not have the topology of $%
{\Bbb R}$, as in local physical theories, but the topology of ${\Bbb R}^{+}$%
. And since in ${\Bbb R}^{+}$ the directions $0\rightarrow \infty $ and $%
\infty \rightarrow 0$ are substantially different, the substantial
difference between the two directions of time would turn out to be imposed
''by hand'' in local theories, which should include the specification of an
absolute origin of time. Moreover, this move would break the Galilean or the
Lorentz invariance of those theories; in particular, the non-homogeneity of
time would lead them to be non-invariant under time-translation and this, in
turn, would amount to resign the local principle of energy conservation.%
\footnote{%
We are grateful to one of the referees for drawing our attention on the
relevance of discussing this point.}

Summing up, local theories do not offer a non-conventional criterion for
distinguishing between the time-symmetric twins and, therefore, between the
two directions of time. When this fact is accepted, general relativity comes
into play and the approach to the problem of the arrow of time turns out to
be global.\bigskip

{\bf 2. Why non-entropic?\smallskip }

When, in the late nineteenth century, Boltzmann developed the probabilistic
version of his theory in response to the objections raised by Loschmidt and
Zermelo (for historical details, see Brush$^{(15)}$), he had to face a new
challenge: how to explain the highly improbable current state of our world.
In order to answer this question, Boltzmann offered the first global
approach to the problem.\footnote{%
Boltzmann$^{(16)}$ wrote: ''{\it The universe, or at least a big part of it
around us, considered as a mechanical system, began in a very improbable
state and it is now also in a very improbable state. Then if we take a
smaller system of bodies, and we isolate it instantaneously from the rest of
the world, in principle this system will be in an improbable state and,
during the period of isolation, it will evolve towards more probable states}%
''.} Since that seminal work, many authors have related the temporal
direction past-to-future to the gradient of the entropy function of the
universe: it has been usually assumed that the fundamental criterion for
distinguishing between the two directions of time is the second law of
thermodynamics (see, for instance, Reichenbach$^{(17)}$, Feynman$^{(18)}$,
Davies$^{(19,20)}$).

The global entropic approach rests on two assumptions: that it is possible
to define entropy for a complete cross-section of the universe, and that
there is an only time for the universe as a whole. However, both assumptions
involve difficulties. In the first place, the definition of entropy in
cosmology is still a very controversial issue: there is no consensus
regarding how to define a global entropy for the universe. In fact, it is
usual to work only with the entropy associated with matter and radiation
because there is not yet a clear idea about how to define the entropy due to
the gravitational field. In the second place, when general relativity comes
into play, time cannot be conceived as a background parameter which, as in
pre-relativistic physics, is used to mark the evolution of the system.
Therefore, the problem of the arrow of time cannot legitimately be posed,
from the beginning, in terms of the entropy gradient of the universe
computed on a background parameter of evolution.

Nevertheless, there is an even stronger argument for giving up the
traditional entropic approach. As it is well known, entropy is a
phenomenological property whose value is compatible with many configurations
of a system. The question is whether there is a more fundamental property of
the universe which allows us to distinguish between both temporal
directions. On the other hand, if the arrow of time reflects a substantial
difference between both directions of time, it is reasonable to consider it
as an intrinsic property of time, or better, of spacetime, and not as a
secondary feature depending on a phenomenological property. For these
reasons we will follow Earman's ''{\it Time Direction Heresy}''$^{(21)}$,
according to which the arrow of time is an intrinsic property of spacetime,
which does not need to be reduced to non-temporal features.

\subsection{Conditions for a global and non-entropic arrow of time}

In general relativity, the universe is a four-dimensional object, physically
described by the geometrical properties of spacetime, embodied in the metric
tensor $g_{\mu \nu }$, and the distribution of matter-energy throughout the
spacetime, embodied in the energy-momentum tensor $T_{\mu \nu }$. Both
properties are physical, and they are related by the Einstein field
equations in such a way that the universe can be physically described in
geometrical terms or in matter-energy terms. We will use a geometrical
language for presenting the conditions for the arrow of time only because it
makes the explanation more intuitive.

As it is well known, many different spacetimes, of extraordinarily varied
topologies, are consistent with the field equations. And some of them have
features that do not admit a unique time for the universe as a whole, or
even the definition of the two directions of time in a global way.
Therefore, the possibility of defining a global arrow of time requires two
conditions that the spacetime must satisfy: time-orientability and existence
of a global time.\bigskip

{\bf 1. Time-orientability}\smallskip

A spacetime $(M,g_{\mu \nu })$ is {\it time-orientable} if there exists a
continuous non-vanishing vector field $\gamma ^{\mu }(x)$ on the manifold $M$
which is everywhere non-spacelike (see Hawking and Ellis$^{(22)}$). By means
of this field, the set of all light semicones of the manifold can be split
into two equivalence classes, $C_{+}$ (semicones containing the vectors of
the field) and $C_{-}$ (semicones non containing the vectors of the field).
It is clear that the names ''$C_{+}$'' and ''$C_{-}$'' are completely
conventional, and can be interchanged as we wish: the only relevant fact is
that, for all the semicones, each one of them belongs to one and only one of
the two equivalence classes. On the contrary, in a non time-orientable
spacetime it is possible to transform a timelike vector into another
timelike vector pointing to the opposite temporal direction by means of a
continuous transport that always keeps non-vanishing timelike vectors
timelike; therefore, the equivalence classes, $C_{+}$ and $C_{-}$ cannot be
defined in an univocal way.\bigskip

{\bf 2. Global time\smallskip }

Time-orientability does not guarantee yet that we can talk of {\it the} time
of the universe: the spacetime may be non globally splittable into spacelike
hypersurfaces such that each one them contains all the events simultaneous
with each other. The {\it stable causality condition} amounts to the
existence of a {\it global time function} on the spacetime (see Hawking and
Ellis$^{(22)}$), that is, a function $t:M\rightarrow {\Bbb R}$ whose
gradient is timelike everywhere. This condition guarantees that the
spacetime can be {\it foliated} into hypersurfaces of simultaneity ($t=const$%
), which can be ordered according to the value of $t$ (see Schutz$^{(23)}$).

\subsection{The definition of the global and non-entropic arrow}

{\bf 1. Time-asymmetry\smallskip }

As Gr\"{u}nbaum$^{(24)}$ correctly points out, the mere ''oppositeness'' of
the two directions of a global time, and even of the two equivalence classes
of semicones, does not provide a non-conventional criterion for
distinguishing the two temporal directions. Such a criterion is given by the
time-asymmetry of the spacetime. A time-orientable spacetime $(M,g_{\mu \nu
})$ with global time $t$ is {\it time-symmetric} with respect to some
spacelike hypersurface $t=\alpha $ if there is a diffeomorphism $d$ of $M$
onto itself which (i) reverses the temporal orientations, (ii) preserves the
metric $g_{\mu \nu }$, and (iii) leaves the hypersurface $t=\alpha $ fixed.
Intuitively, this means that the spacelike hypersurface $t=\alpha $ splits
the spacetime into two ''halves'', one the temporal mirror image of the
other. On the contrary, in a time-asymmetric spacetime there is no spacelike
hypersurface $t=\alpha $ from which the spacetime looks the same in both
temporal directions: the properties in one direction are different than the
properties in the other direction, and this fact is expressed by the metric $%
g_{\mu \nu }$. But, according to the Einstein field equations, this also
means that the matter-energy of the universe is asymmetrically distributed
along the global time, and this is expressed by the energy-momentum tensor $%
T_{\mu \nu }$. Therefore, no matter which spacelike hypersurface is used to
split a time-asymmetric spacetime into to ''halves'', the physical
(geometrical or matter-energy) properties of both halves are substantially
different, and such a difference establishes a substantial distinction
between the two directions of time.

Now we can assign different names to the substantially different temporal
directions on the basis of that difference. For instance, we can call one of
the directions of the global time $t$ ''positive'' and the class of
semicones containing vectors pointing to positive $t$ ''$C_{+}$'', and the
other direction of $t$ ''negative'' and the corresponding class ''$C_{-}$''.
Of course, the particular names chosen are absolutely conventional, we can
use the opposite convention, or even other names (''$A$'' and ''$B$'',
''black'' and ''white'', or ''Alice'' and ''Bob''): the only relevant fact
is that both directions of time are substantially different to each other,
and the different names assigned to them express such a substantial
difference.\bigskip

{\bf 2. The meaning of time-reversal invariance in general
relativity\smallskip }

The metric $g_{\mu \nu }$ and the energy-momentum tensor $T_{\mu \nu }$ of a
particular spacetime are, of course, a solution of the Einstein field
equations which, being fundamental laws, are time-reversal invariant. Then,
there exists another solution given by ${\cal T\,}g_{\mu \nu }$ and ${\cal %
T\,}T_{\mu \nu }$, the time-symmetric twin of the previous one. So, the
ghost of symmetry threatens again: it seems that we are committed to
supplying a non-conventional criterion for picking out one of both
solutions, one the temporal mirror image of the other. However, in this case
the threat is not as serious as it seems.

As it is well known, time-reversal is a symmetry transformation. Under the
active interpretation, a symmetry transformation corresponds to a change
from one system to another; under the passive interpretation, a symmetry
transformation consists in a change of the point of view from which the
system is described. The traditional position about symmetries assumes that,
in the case of discrete transformations as time-reversal or spatial
reflection, only the active interpretation makes sense: an idealized
observer can rotate himself in space in correspondence with the given
spatial rotation, but it is impossible to ''rotate in time'' (see Sklar$%
^{(25)}$). Of course, this is true when the idealized observer is immersed
in the same spacetime as the observed system. But when the system is the
universe as a whole, we cannot change our spatial perspective with respect
to the universe: it is as impossible to rotate in space as to rotate in
time. However, this does not mean that the active interpretation is the
correct one: the idea of two identical universes, one translated in space or
in time regarding the other, has no meaning. This shows that both
interpretations, when applied to the universe as a whole, collapse into
conceptual nonsense (for a full argument, see Castagnino, Lombardi and Lara$%
^{(1)}$).

In fact, in cosmology symmetry transformations are neither given an active
nor a passive interpretation. Two models $(M,g_{\mu \nu })$\ and $(M^{\prime
},g_{\mu \nu }^{\prime })$ of the Einstein equations are taken to be
equivalent if they are {\it isometric}, that is, if there is a
diffeomorphism $\theta :M\rightarrow M^{\prime }$\ which carries the metric $%
g_{\mu \nu }$\ into the metric $g_{\mu \nu }^{\prime }$\ (see Hawking and
Ellis$^{(22)}$). Since symmetry transformations are isometries, two models
related by a symmetry transformation (in particular, time-reversal) are
considered equivalent descriptions of one of the same spacetime. Therefore,
by contrast with local theories, when the object described is the universe
as a whole, it is not necessary to supply a non-conventional criterion for
selecting one solution of the pair of time-symmetric twins. This fundamental
difference between general relativity and the local theories of physics is
what allows the global approach to the problem of the arrow of time to
provide a solution that cannot be offered by local approaches.\bigskip

{\bf 3. The generic character of the global and non-entropic arrow\smallskip
}

As we have seen, the global entropic approach explains the arrow of time in
terms of the increasing entropy function of the universe: as a consequence,
this position has to posit a low-entropy initial state from which entropy
increases. Then, the problem of the arrow of time is pushed back to the
question of why the initial state of the universe has low entropy. But a
low-entropy initial state is extraordinarily improbable in the collection of
all possible initial states. Therefore, the global entropic approach is
committed to supply an answer to the problem of explaining such an
improbable initial condition (see, for instance, the arguments of Davies$%
^{(19,20)}$ and of Penrose and Percival$^{(26)}$; see also the well known
criticisms directed by Price$^{(13)}$ to the global entropic approach).

In our global and non-entropic approach there are not improbable conditions
that require to be accounted for. On the contrary, in previous papers
(Castagnino and Lombardi$^{(4,5)}$) we have proved that the subset of
time-symmetric spacetimes has measure zero in (or is a proper subspace of)
the set of all possible spacetimes admissible in general relativity. This
result can be intuitively understood on the basis of the evident fact that
symmetry is a very specific property, whereas asymmetry is greatly generic.
Therefore, in the collection of all the physically possible spacetimes,
those endowed with a global and non-entropic arrow of time are
overwhelmingly probable: the non existence of the arrow of time is what
requires an extraordinarily fine-tuning of all the variables of the universe.

These arguments, based on theoretical results, are relevant to the problem
of finding a substantial difference between the two directions of time
grounded on physical theories. Of course, theories are undetermined by
empirical evidence, and this underdetermination is even stronger in
cosmology, where the observability horizons of the universe introduce
theoretical limits to our access to empirical data. In fact, on the basis of
the features of the unobservable regions of the universe, it may be the case
that our spacetime be time-symmetric, or lacking a global time, or even non
time-orientable. Of course, this does not undermine the overwhelmingly low
probability of time-symmetry. But since probability zero does not amount to
impossibility, we cannot exclude the case that we live in a time-symmetric,
or even in a non time-orientable universe. In that case, the global and
non-entropic arrow of time would not exist and, therefore, the explanation
of the local time-asymmetries to be presented in the next sections would not
apply. It is quite clear that this case cannot be excluded on logical nor on
theoretical grounds. However, the coherence of our overall explanation of
the arrow of time and of the local irreversible phenomena, whose theoretical
account we were looking for, counts for its plausibility. On the other hand,
it is difficult to see what theoretical or empirical reasons could be used
to argue for the fact that we live in a universe lacking a global arrow of
time; on the contrary, the cosmological models accepted in present-day
cosmology as the best representations of our actual universe (Big Bang-Big
Rip FRW models) are clearly time-asymmetric (see Caldwell {\it et al}.$%
^{(27)}$). As always in physics and, in general, in science, there are not
irrefutable explanations. In particular, any statistical argument admits
probability zero exceptions. Nevertheless, we can make reasonable decisions
about accepting or rejecting a particular explanation on the basis of its
fruitfulness for explaining empirical evidence and its coherence with the
knowledge at our disposal.\footnote{%
We are grateful to one of the referees for stressing the need to discuss
this point.}

\subsection{Transferring the global arrow to local contexts}

{\bf 1. From time-asymmetry to energy flow\smallskip }

As we have seen, the time-asymmetry of spacetime establishes a substantial
difference between the two directions of time. This time-asymmetry is a
physical property of the spacetime that can be equivalently expressed in
geometrical terms ($g_{\mu \nu }$) or in terms of the matter-energy
distribution ($T_{\mu \nu }$). However, in none of both descriptions it can
be introduced in local theories which, in principle, do not contain the
concepts of metric or of energy-momentum tensor. For this reason, if we want
to transfer the global arrow of time to local contexts, we have to translate
the time-asymmetry embodied in $g_{\mu \nu }$ and $T_{\mu \nu }$ into a
feature that can be expressed by the concepts of local theories. This goal
can be achieved by expressing the energy-momentum tensor in terms of the
four-dimensional energy flow.

As it is well known, in the energy-momentum tensor $T_{\mu \nu }$, the
component $T_{00}$ represents the matter-energy density and the component $%
T_{0i}$, with $i=1$ to $3$, represents the spatial energy flow. Thus, $%
T_{0\alpha }$ can be viewed as a spatio-temporal matter-energy flow that
embodies, not only the flow of matter-energy in space but also its flow in
time; let us call it ''energy flow'' for simplicity. In turn, $T_{\mu \nu }$
satisfies the dominant energy condition if, in any orthonormal basis, $%
T_{00}\geq \left| T_{\alpha \beta }\right| $, for each $\alpha ,\beta =0$ to
$3$. This is a very weak condition, since it holds for almost all known
forms of matter-energy (see Hawking and Ellis$^{(22)}$) and, then, it can be
considered that $T_{\mu \nu }$ satisfies it in almost all the points of the
spacetime. The dominant energy condition means that, for any local observer,
the matter-energy density $T_{00}$ is {\it non-negative} and the energy flow
$T_{0\alpha }$ is {\it non-spacelike}. Therefore, in all the points of the
spacetime where the condition holds, $T_{00}$ points to the same temporal
direction and, since it gives the temporal direction of $T_{0\alpha }$, the
energy flow is endowed with the same feature.

The first point to stress here is that the dominant energy condition is
immune to time-reversal. In fact, if we apply the time-reversal operator $%
{\cal T}$ to $T_{\mu \nu }$, we obtain
\begin{eqnarray}
{\cal T\,}T_{00} &=&T_{00}\qquad \qquad {\cal T\,}T_{i0}=-T_{i0}  \nonumber
\\
{\cal T\,}T_{0i} &=&-T_{0i}\qquad \qquad {\cal T\,}T_{ij}=T_{ij}  \label{19}
\end{eqnarray}
Therefore, time-reversal does not change the sign of $T_{00}$ and, as a
consequence, if the condition is satisfied by $T_{\mu \nu }$, it is also
satisfied by ${\cal T\,}T_{\mu \nu }$.

The second point that has to be emphasized is that, without a substantial
difference between the two temporal directions, the term ''positive''
applied to $T_{00}$ by the dominant energy condition is merely conventional.
The relevant content of the condition is that the energy flow $T_{0\alpha }$
is {\it non-spacelike} (that is, the matter-energy does not flow faster than
light), and that it points to the same temporal direction in all the points
where the condition is satisfied. Therefore, we can choose any temporal
direction as ''positive'', and the condition will preserve its conceptual
meaning.

A third point that deserves to be stressed is the application of the
condition to spacetimes where $T_{00}$ points to the ''opposite'' temporal
direction than in the original spacetime. Let us consider the two spacetimes
with $T_{\mu \nu }^{\prime }=-T_{\mu \nu }$ and $T_{\mu \nu }^{\prime \prime
}=-{\cal T\,}T_{\mu \nu }$:
\begin{eqnarray}
T_{00}^{\prime } &=&-T_{00}\qquad \qquad T_{00}^{\prime \prime }=-T_{00}
\nonumber \\
T_{0i}^{\prime } &=&-T_{0i}\qquad \qquad T_{0i}^{\prime \prime }=T_{0i}
\label{20}
\end{eqnarray}
It is clear that, if the dominant energy condition is satisfied by $T_{\mu
\nu }$, it is also satisfied by $T_{\mu \nu }^{\prime }$ and $T_{\mu \nu
}^{\prime \prime }$, with the only change in the conventional decision about
the ''positive'' time direction. This is not surprising since the models
with $T_{\mu \nu }$, $T_{\mu \nu }^{\prime }$ and $T_{\mu \nu }^{\prime
\prime }$ are isometric and, therefore, they are equivalent descriptions of
one and the same universe: to say that the temporal direction in one of them
is ''opposite'' to the time direction in the other is senseless (see
Subsection III.C.2).

Once these delicate points have been understood,\footnote{%
We are grateful to one of the referees for drawing our attention on the need
to explain these subtle points.} it turns out to be clear that the
non-conventional content of the dominant energy condition is not the
substantial identification of a temporal direction as the direction of the
energy flow, but the coordination of the temporal direction of that flow in
all the points of the spacetime where the condition holds. In other words,
if $T_{\mu \nu }$ satisfies the condition in all (or almost all) the points
of the spacetime, we can guarantee that the energy flow is always contained
in the semicones belonging to only one of the two classes, $C_{+}$ or $C_{-}$%
, arising from the partition introduced by time-orientability. At this
point, somebody might suppose that the satisfaction of the condition, by
itself, solves the problem of the arrow of time: ''if the energy flow points
to the same temporal direction all over the spacetime, let us define that
direction as the future and that's all, we don't need time-asymmetry''. But
this conclusion forgets the conventionality of the direction selected as
''positive'' for the energy flow. In fact, even if the energy flow is always
contained in the semicones belonging to, say, $C_{+}$, in a time-symmetric
spacetime the difference between $C_{+}$ and $C_{-}$ is merely conventional.
Only when we can substantially distinguish between the two temporal
directions in terms of the time-asymmetry of the spacetime, we can use the
energy flow pointing to the same direction all over the spacetime for
expressing that substantial difference. In short, the arrow of time is {\it %
defined} by the time-asymmetry of the spacetime, and {\it expressed} by the
energy flow.

Up to now we have not used the words ''past'' and ''future''. It is clear
that, since they are only labels, their application is conventional. Now we
can introduce the usual convention in physics, which consists in calling the
temporal direction of $T_{00}$ and, therefore, also the temporal direction
of the energy flow $T_{0\alpha }$ ''positive direction'' or ''future'';
then, it is said that, under the dominant energy condition, $T_{0\alpha }$
is always contained in the semicones belonging to the positive class $C_{+}$%
. With this convention we can say that the energy flows towards the future
for any observer and in any point of the spacetime.\footnote{%
If we could compute the global entropy of the universe, with this convention
we could say that the entropy increases towards the future. But this would
require that the technical difficulties for defining such an entropy were
overcome (see Subsection III.A.2).} Of course, we could have used the
opposite convention and have said that the energy always flows towards the
past. But, in any case, no matter which terminological decision we make,
past is substantially different than future, and the arrow of time consists
precisely in such a substantial difference grounded on the time-asymmetry of
spacetime.\bigskip

{\bf 2. Breaking local time-symmetries\smallskip }

As we have said, any time-reversal invariant equation leads to a pair of
time-symmetric twins, that is, two solutions symmetrically related by the
time-reversal transformation: each twin, which in some cases represents an
irreversible evolution, is the temporal mirror image of the other twin. From
the viewpoint of the local theory that produces the time-symmetric twins,
the difference between them is only conventional: both twins are
nomologically possible with respect to the theory. And, as the Eherenfests
pointed out, this is also true for entropy when computed in terms of a
time-reversal invariant fundamental law. The traditional arguments for
discarding one of the twins and retaining the other invoke temporally
asymmetric notions which are not justified in the context of the local
theory. For instance, the retarded nature of radiation is usually explained
by means of {\it de facto} arguments referred to initial conditions:
advanced solutions of wave equations correspond to converging waves that
require a highly improbable ''conspiracy'', that is, a ''miraculous''
cooperative emitting behavior of distant regions of space at the temporal
origin of the process (see Sachs$^{(12)}$).

It seems quite clear that this kind of arguments, even if admissible in the
discussions about irreversibility, are not legitimate in the context of the
problem of the arrow of time, to the extent that they put the arrow ''by
hand'' by presupposing the difference between the two directions of time
from the very beginning. In other words, they violate the ''nowhen''
requirement of adopting an atemporal perspective purged of our temporal
intuitions and our time-asymmetric observations, like those related with the
asymmetry between past and future or between initial and final conditions.
Therefore, from an atemporal standpoint, the challenge consists in supplying
a non-conventional criterion, {\it based only on theoretical arguments}, for
distinguishing between the two members of the pair of twins. The desired
criterion, that can be legitimately supplied neither by the local theory nor
by our time-asymmetric experiences, can be grounded on global considerations.

If we adopt the usual terminological convention according to which the
future is the temporal direction of the energy flow and $C_{+}(x)$ denotes a
future semicone, then the energy flow is contained in the future semicone $%
C_{+}(x)$ for any point $x$ of the spacetime. On the other hand, in any pair
of time-symmetric twins, the members of the pair involve energy flows
pointing to opposite temporal directions, with no non-conventional criterion
for distinguishing between them. But once {\it we have established the
substantial difference between past and future on global grounds} and have
decided that energy flows towards the future, we have a substantial
criterion for discarding one of the twins and retaining the other as
representing the relevant solution of the time-reversal invariant law. For
instance, given the usual conventions, in the case of electromagnetism only
retarded solutions are retained, since they describe states that carry
energy towards the future.

Another relevant example is the creation and decaying of unstable states.
From an equilibrium state $\rho _{*}$, an unstable non-equilibrium state $%
\rho (t=0)$ is created by an antidissipative process with evolution factor $%
e^{\gamma t\text{ }}$and $\sigma <0$. This unstable non-equilibrium state $%
\rho (t=0)$ decays towards an equilibrium state $\rho _{*}$ through a
dissipative process with evolution factor $e^{-\gamma t\text{ }}$ and $%
\sigma >0$. When considered locally, the pairs of twins (antidissipativea
and dissipative, $\sigma <0$ and $\sigma >0$) are only conventionally
different. However, since past is substantially different than future and,
according to the usual convention, the energy flow always goes from past to
future, the unstable states are always created by energy pumped from the
energy flow coming from the past, while unstable states decay returning this
energy to the energy flow pointing towards the future. Therefore, the energy
flow introduces a substantial difference between the two members of each
pair.

In the next sections, we will analyze the breaking of the symmetry
introduced by the energy flow in different local laws, coming from
fundamental theories and from phenomenological theories; in this last case,
the first step will be to bring into the light the second twin usually
hidden in the formalism.

\section{Fundamental theories}

\subsection{Quantum mechanics}

The so-called irreversible quantum mechanics is based on the use of rigged
Hilbert spaces, due to the ability of this formalism to model irreversible
physical phenomena such as exponential decay or scattering processes (see
Bohm and Gadella$^{(28)}$). The general strategy consists in introducing two
subspaces, $\Phi _{-}$ and $\Phi _{+}$, of the Hilbert space ${\cal H}$. The
vectors $|\phi \rangle $ of the subspaces $\Phi _{-}$ and $\Phi _{+}$ are
characterized by the fact that their projections $\langle \omega |\phi
\rangle $ on the eigenstates $\omega $ of the energy are functions of the
Hardy class from above and from below respectively. These subspaces yield
two rigged Hilbert spaces:
\begin{equation}
\Phi _{-}\subset {\cal H\subset }\Phi _{-}^{\times }\qquad \qquad \Phi
_{+}\subset {\cal H\subset }\Phi _{+}^{\times }  \label{21}
\end{equation}
where $\Phi _{-}^{\times }$ and $\Phi _{+}^{\times }$ are the anti-dual of
the spaces $\Phi _{-}$ and $\Phi _{+}$ respectively.

It is quite clear that, up to this point, this general strategy amounts to
obtain two time-symmetric structures from a time-reversal invariant theory.
In fact, quantum mechanics formulated on a Hilbert space ${\cal H}$ is
time-reversal invariant since
\begin{equation}
{\bf T}{\cal \,H=H}  \label{22}
\end{equation}
where ${\bf T}$ is the antilinear and antiunitary time-reversal operator
(see Section II.A). But if quantum mechanics is formulated on spaces $\Phi
_{\pm }$, it turns out to be a non time-reversal invariant theory since
\begin{equation}
{\bf T\,}\Phi _{\pm }=\Phi _{\mp }  \label{23}
\end{equation}
Moreover, in the analytical continuation of the energy spectrum of the
system's Hamiltonian, there exists at least a pair of complex conjugate
poles, one in the lower half-plane and the other in the upper half-plane of
the complex plane. Such poles correspond to a pair of Gamov vectors:
\begin{equation}
|\Psi _{G}^{-}\rangle \in \Phi _{-}^{\times }\qquad \qquad |\Psi
_{G}^{+}\rangle \in \Phi _{+}^{\times }  \label{24}
\end{equation}
These vectors are proposed to describe irreversible processes and are taken
to be the representation of the exponentially growing and decaying part of
resonant unstable states, respectively. But the symmetric position of the
poles with respect to the real axis in the complex plane is a clear
indication of the fact that Gamov vectors are a case of time-symmetric twins.

In his detailed description of scattering processes, Arno Bohm breaks the
symmetry between the twins by appealing to the so-called
''preparation-registration arrow of time'', expressed by the slogan ''{\it %
No registration before preparation}'' (see Bohm, Antoniou and Kielanowski$%
^{(29,30)}$). The key idea behind this proposal is that observable
properties of a state cannot be measured until the state acting as a bearer
of these properties has been prepared. For instance, in a scattering
process, it makes no sense to measure the scattering angle until a state is
prepared by an accelerator. On this basis, Bohm proposes the following {\it %
interpretational postulate}: the vectors $|\varphi \rangle \in \Phi _{-}$
represent the states of the system and the vectors $|\psi \rangle \in \Phi
_{+}$ represent the observables of the system in the sense that observables
are obtained as $O=|\psi \rangle \langle \psi |$. The time $t=0$ is
considered as the time at which preparation ends and detection begins. The
preparation-registration arrow imposes the requirement that the energy
distribution produced by the accelerator, represented by $\langle \omega
|\varphi \rangle $, be zero for $t>0$, and the energy distribution of the
detected state, represented by $\langle \omega |\psi \rangle $, be zero for $%
t<0$:

\begin{eqnarray}
|\varphi \rangle  &\in &\Phi _{-}\qquad \qquad \langle \omega |\varphi
\rangle =0\quad \text{for }t>0  \nonumber \\
|\psi \rangle  &\in &\Phi _{+}\qquad \qquad \langle \omega |\psi \rangle
=0\quad \text{for }t<0  \label{25}
\end{eqnarray}
The time evolution, traditionally represented by the group $U_{t}$ on the
Hilbert space ${\cal H}$, is here represented by two semigroups $U_{t}^{-}$
and $U_{t}^{+}$: (i) $U_{t}^{-}$ is $U_{t}$ restricted to $\Phi _{-}$ and,
then, it is valid only for $t>0$, and (ii) $U_{t}^{+}$ is $U_{t}$ restricted
to $\Phi _{+}$ and, then, it is valid only for $t<0$ (see Bohm {\it et al.}$%
^{(31)}$, Bohm and Wickramasekara$^{(32)}$):
\begin{eqnarray}
U_{t}^{-} &:&\Phi _{-}\rightarrow \Phi _{-}\qquad |\varphi (t)\rangle
=U_{t}^{-}|\varphi _{0}\rangle \quad \text{for }t>0  \nonumber \\
U_{t}^{+} &:&\Phi _{+}\rightarrow \Phi _{+}\qquad |\psi (t)\rangle
=U_{t}^{+}|\psi _{0}\rangle \quad \text{for }t<0  \label{30}
\end{eqnarray}
As a consequence, the two Gamov vectors $|\Psi _{G}^{-}\rangle \in \Phi
_{-}^{\times }$ and $|\Psi _{G}^{+}\rangle \in \Phi _{+}^{\times }$ turn out
to be representations of growing and decaying states respectively: the
evolution of the {\it growing Gamov vector} $|\Psi _{G}^{-}\rangle $ can be
defined only for $t<0$, and the evolution of the {\it decaying Gamov vector}
$|\Psi _{G}^{+}\rangle $ can be defined only for $t>0$.

As we can see, Bohm's approach breaks the symmetry between the
time-symmetric twins by means of an interpretational postulate based on the
preparation-registration arrow. Of course, this strategy supplies a solution
to the problem of irreversibility to the extent that it permits the
representation of irreversible growing and decaying processes. However, it
does not offer a theoretical way out of the problem of the arrow of time,
since the preparation-registration arrow is introduced as a postulate of the
theory. In fact, such an arrow presupposes the distinction between past and
future from the very beginning: in the past ($t<0$) the system is prepared,
in the future ($t>0$) the system is measured, and $|\Psi _{G}^{-}\rangle $
and $|\Psi _{G}^{+}\rangle $ represent the corresponding growing and
decaying processes respectively, both evolving toward the future. It is
clear that such a postulate is based on our pretheoretical assumption that
preparation is temporally previous than registration. But, from an atemporal
viewpoint, we could reverse that interpretational postulate: we could
consider that $\Phi _{+}$ is the space of states and $\Phi _{-}$ is the
space of vectors by means of which the observables are obtained; in this
case we would obtain the temporal mirror image of the original theory, where
$|\Psi _{G}^{+}\rangle $ and $|\Psi _{G}^{-}\rangle $ represent growing and
decaying states respectively, both evolving toward the past. In other words,
the two possible interpretational postulates restore the time-symmetry since
they lead to two non time-reversal invariant theoretical structures, one the
temporal mirror image of the other. Bohm's strategy of choosing the future
directed version of the postulate introduces the arrow of time only on the
basis of pretheoretical considerations (for a detailed discussion, see
Castagnino, Gadella and Lombardi$^{(33,34)}$) .

Nevertheless, it is possible to find a theoretical justification for the
preparation-registration arrow and, as a consequence, for choosing the
future directed version of the interpretational postulate: the breaking of
the symmetry between the time-symmetric twins is supplied by the energy flow
represented by $T_{0\alpha }$. The preparation of the states $|\varphi
\rangle \in \Phi _{-}$ acting as bearers of properties requires energy
coming from other processes. Since the energy flow comes from the past and
goes to the future, the states are prepared by means of energy coming from
the past, that is, from previous processes; the growing Gamov vector $|\Psi
_{G}^{-}\rangle $ represents precisely the growing process occurring at $t<0$
, which absorbs the energy coming from the past. On the other hand, the
registration of the observable properties represented by $|\psi \rangle \in
\Phi _{+}$ provides energy to other processes. Again, since the energy flow
comes from the past and goes to the future, the measurement of observables
emits energy toward the future, that is, provides energy to latter
processes; the decaying Gamov vector $|\Psi _{G}^{+}\rangle $ represents
precisely the decaying process occurring at $t>0$, which emits energy toward
the future. Summing up, the energy flow coming from the past and directed
towards the future supplies the criterion for selecting the future directed
version of Bohm's interpretational postulate, and turns quantum mechanics
into a non time-reversal invariant theory without the addition of
pretheoretical assumptions.

\subsection{Quantum field theory}

It is quite clear that the two classes of light semicones $C_{+}$ and $C_{-}$
are a pair of time-symmetric twins. Quantum field theory breaks the symmetry
of this pair from the very beginning, by introducing non time-reversal
invariance as a primitive assumption. In this section we will explain how
this symmetry-breaking can be derived from the global time-asymmetry of the
universe.\bigskip

{\bf 1. The non time-reversal invariance of axiomatic QFT\smallskip }

In any of its versions, axiomatic QFT includes a non time-reversal invariant
postulate (see Bogoliubov {\it et al.}$^{(35)}$, Roman$^{(36)}$, and also
Haag$^{(37)}$, where it is called Postulate III), which states that the
spectrum of the energy-momentum operator $P^{\mu }$ is confined to a future
light semicone, that is, its eigenvalues $p^{\mu }$ satisfy
\begin{equation}
p^{2}\geq 0\qquad \qquad p^{0}\geq 0  \label{31}
\end{equation}
This postulate says that, when we measure the observable $P^{\mu }$, we
obtain a {\it non-spacelike} {\it classical} $p^{\mu }$ {\it contained in a
future semicone}, that is, a semicone belonging to $C_{+}$.

It is clear that condition $p^{0}\geq 0$ selects one of the elements of the
pair of time-symmetric twins $C_{+}$ and $C_{-}$ or, in other words, of the
pair $p^{0}\geq 0$ and $p^{0}\leq 0$ that would arise from the theory in the
absence of the time-reversal invariance breaking postulate. By means of this
postulate, QFT becomes a non time-reversal invariant theory. In turn, since
QFT, being both quantum and relativistic, can be considered one of the most
basic theories of physics, the choice introduced by condition $p^{0}\geq 0$
is transferred to the rest of physical theories. But such a choice is
established from the very beginning, as an unjustified assumption. The
challenge is, then, to {\it justify } the non time-reversal invariant
postulate by means of independent theoretical arguments.

Let us recall that, in the energy-momentum tensor, $T_{0\alpha }$ represents
the spatio-temporal matter-energy flow and $T_{\alpha 0}$ represents the
linear momentum density. Since $T_{\mu \nu }$ is a symmetric tensor, $T_{\mu
\nu }=T_{\nu \mu }$ and, therefore, $T_{0\alpha }=T_{\alpha 0}$; in other
words, the matter-energy flow is equal to the linear momentum density. This
means that, if $T_{0\alpha }$ can be used to express the global arrow of
time under the dominant energy condition, this is also the case for the
linear momentum density $T_{\alpha 0}$. But it is precisely the linear
momentum density $T_{\alpha 0}$ the magnitude corresponding to the classical
$p^{\mu }$ of QFT; thus, at each point $x$ of the spacetime, $T_{\alpha
0}(x)\in C_{+}(x)\Longrightarrow p^{\mu }\sim T_{\alpha 0}(x)\in C_{+}(x)$.

In conclusion, the fact that{\bf \ }$p^{\mu }$ at each point $x$ of the
local context and, therefore, for every classical particle, must be
contained in the future light semicone $C_{+}(x)$ turns out to be a
consequence of the global time-asymmetry of the spacetime when the dominant
energy condition holds everywhere. In other words, the non time-reversal
invariant postulate can be justified on global grounds instead of being
imposed as a starting point of the axiomatic version of QFT.\bigskip

{\bf 2. The non time-reversal invariance of ordinary QFT\smallskip }

In the ordinary version of QFT, the classification of one-particle states
according to their transformation under the Lorentz group leads to six
classes of four-momenta. Among these classes, it is considered that only
three have physical meaning: these are precisely the cases that agree with
the non time-reversal invariant postulate of the axiomatic version of QFT.
In other words, the symmetry group of QFT is the orthochronous group (see
Weinberg$^{(38)}$), where space-reversal ${\cal P}$ but not time-reversal $%
{\cal T}$ is included. This is another way of expressing the non
time-reversal invariance of QFT. In this case, the non time-reversal
invariance is introduced not by means of a postulate, but on the basis of
empirical arguments that make physically meaningless certain classes of
four-momenta. However, to the extent that special relativity and standard
quantum mechanics are time-reversal invariant theories, those arguments give
no theoretically grounded justification for such a breaking of time-reversal
invariance. Nevertheless, as we have seen in the previous subsection, this
justification can be given on global grounds.

Let us make the point in different terms. The quantum field correlates of $%
{\cal P}$ and ${\cal T}$, ${\bf P}$ and ${\bf T}$, are defined as
\begin{equation}
{\bf P}iP^{\nu }{\bf P}^{-1}=i{\cal P}_{\mu }^{\nu }P^{\mu }\qquad \qquad
{\bf T}iP^{\nu }{\bf T}^{-1}=i{\cal T}_{\mu }^{\nu }P^{\mu }  \label{32}
\end{equation}
where ${\bf P}$ is a linear and unitary operator and ${\bf T}$ is an
antilinear and antiunitary operator (see Section II.A). In fact, if ${\bf T}$
were linear and unitary, we could simply cancel the $i$'s and, then, from
eq. (\ref{32}), ${\bf T}P^{0}{\bf T}^{-1}=-P^{0}$: the action of the
operator ${\bf T}$ on the operator $P^{0}$ would invert the sign of $P^{0}$,
with the consequence that the spectrum of the inverted energy-momentum
operator would be contained in a past light semicone. Precisely, for $\nu =0$%
, $P^{\nu }=H$, where $H$ is the energy operator; then, if ${\bf T}$ were
linear and unitary, ${\bf T}H{\bf T}^{-1}=-H$ (in contradiction with eq. (%
\ref{3})) with the consequence that, for any state of energy $E$ there would
be another state of energy $-E$. The antilinearity and the antiunitarity of $%
{\bf T}$ avoid these ''anomalous'' situations, in agreement with the
conditions imposed by the non time-reversal invariant postulate and, at the
same time, make QFT non time-reversal invariant. Once again, there are good
empirical reasons for making ${\bf T}$ antilinear and antiunitary, but not
theoretical justification for such a move.

Summing up, in ordinary QFT it is always necessary to make a decision about
the time direction of the spectrum of the energy-momentum operator $P^{\mu }$%
. The point that we want to stress here is that, either in the case of the
non time-reversal invariant postulate of the axiomatic version of QFT or in
the case of the usual version of QFT, the decision can be justified on
global grounds, as a consequence of the time-asymmetry of the
spacetime.\bigskip

{\bf 3. Weak interactions\smallskip }

Finally, it is worth reflecting on the role of weak interactions in the
problem of the arrow of time. The CPT theorem states that ${\bf CPT}$ is the
only combination of charge-conjugation ${\bf C}$, parity-reflection ${\bf P}$
and time-reversal ${\bf T}$ which is a symmetry of QFT. In fact, it is well
known that weak interactions break the ${\bf T}$ of the CPT theorem.
According to a common opinion, it is precisely this empirical fact the clue
for the solution of the problem of the arrow of time: since the ${\bf T}$
symmetry is violated by weak interactions, they introduce a non-conventional
distinction between the two directions of time (see Visser$^{(39)}$). The
question is: Is the breaking of ${\bf T}$ what distinguishes both directions
of time in QFT? As we have seen, the operator ${\bf T}$ was designed
precisely to avoid that certain tetra-magnitudes, such as the linear
momentum $p^{\mu }$, have the ''anomalous'' feature of being contained in a
past light semicone: the action of the operator ${\bf T}$ onto the
energy-momentum operator $P^{\mu }$ preserves the time direction of $P^{\mu
} $ and, therefore, of its eigenvalues. It is this fundamental fact what
makes QFT non time-reversal invariant, and not the incidental violation of $%
{\bf T} $ by weak interactions. This non time-reversal invariance of QFT,
based on the peculiar features of the operator ${\bf T}$, distinguishes by
itself between the two directions of time, with no need of weak interactions
(see discussion in Castagnino and Lombardi$^{(4)}$). In other words, even if
weak interactions did not exist, QFT would be a non time-reversal invariant
theory which would distinguish between the two directions of time.\footnote{%
Of course, this leaves open a different problem: to explain why, among all
the elementary interactions, only weak interactions break time-reversal
invariance. It has be suggested that weak interactions define the arrow of
time, but it is not clear at all how this microscopic phenomenon could break
the time-asymmetry of the twins arising at the macroscopic level.} The real
problem is, then, to justify the non time-reversal invariance of a theory
which is presented as a synthesis of two time-reversal invariant theories
such as special relativity and quantum mechanics. But this problem is
completely independent of the existence of weak interactions and the
breaking of ${\bf T}$ introduced by them. Summing up, weak interactions do
not play a role as relevant in the problem of the arrow of time as it is
usually supposed.

\subsection{Feynman graphs and quantum measurements}

{\bf 1. Feynman graphs\smallskip }

\begin{figure}[h]
\begin{center}
\includegraphics[width=6cm,angle=0]{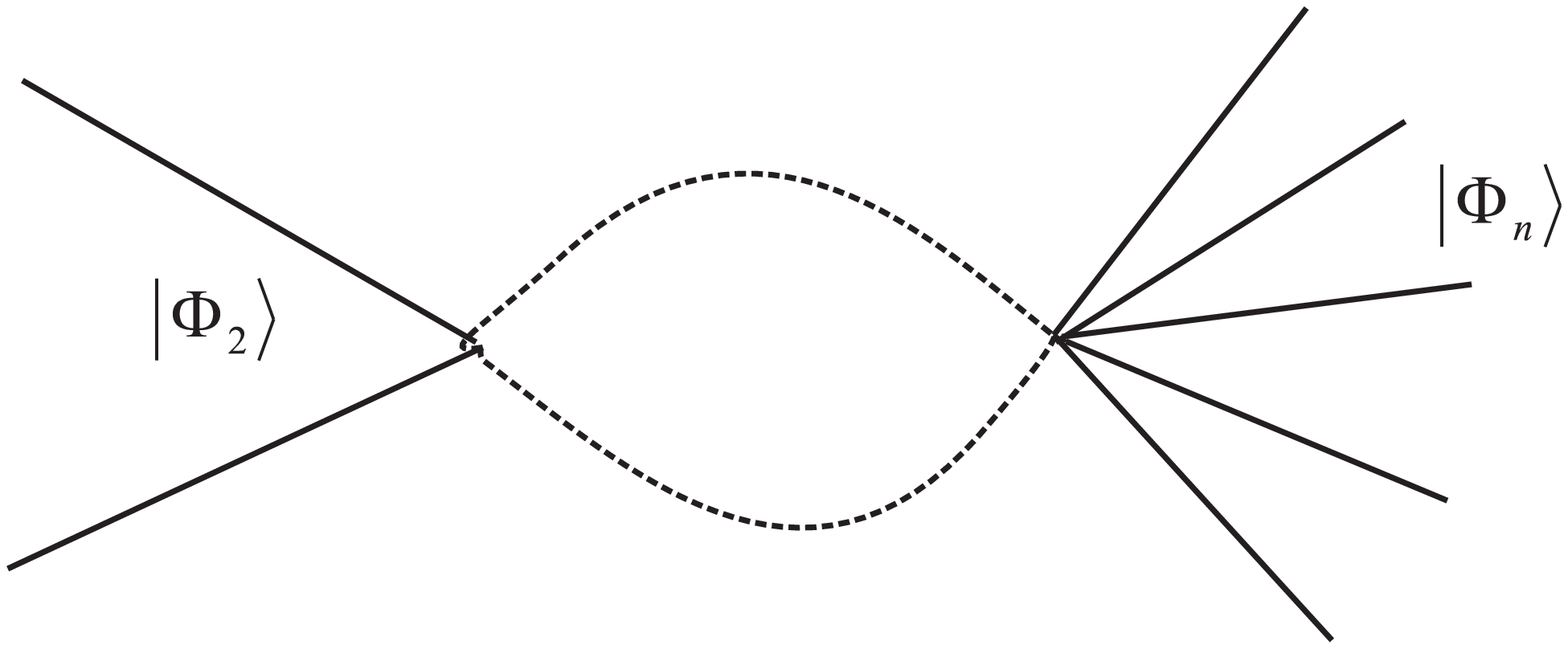} \\
\includegraphics[width=5cm,angle=0]{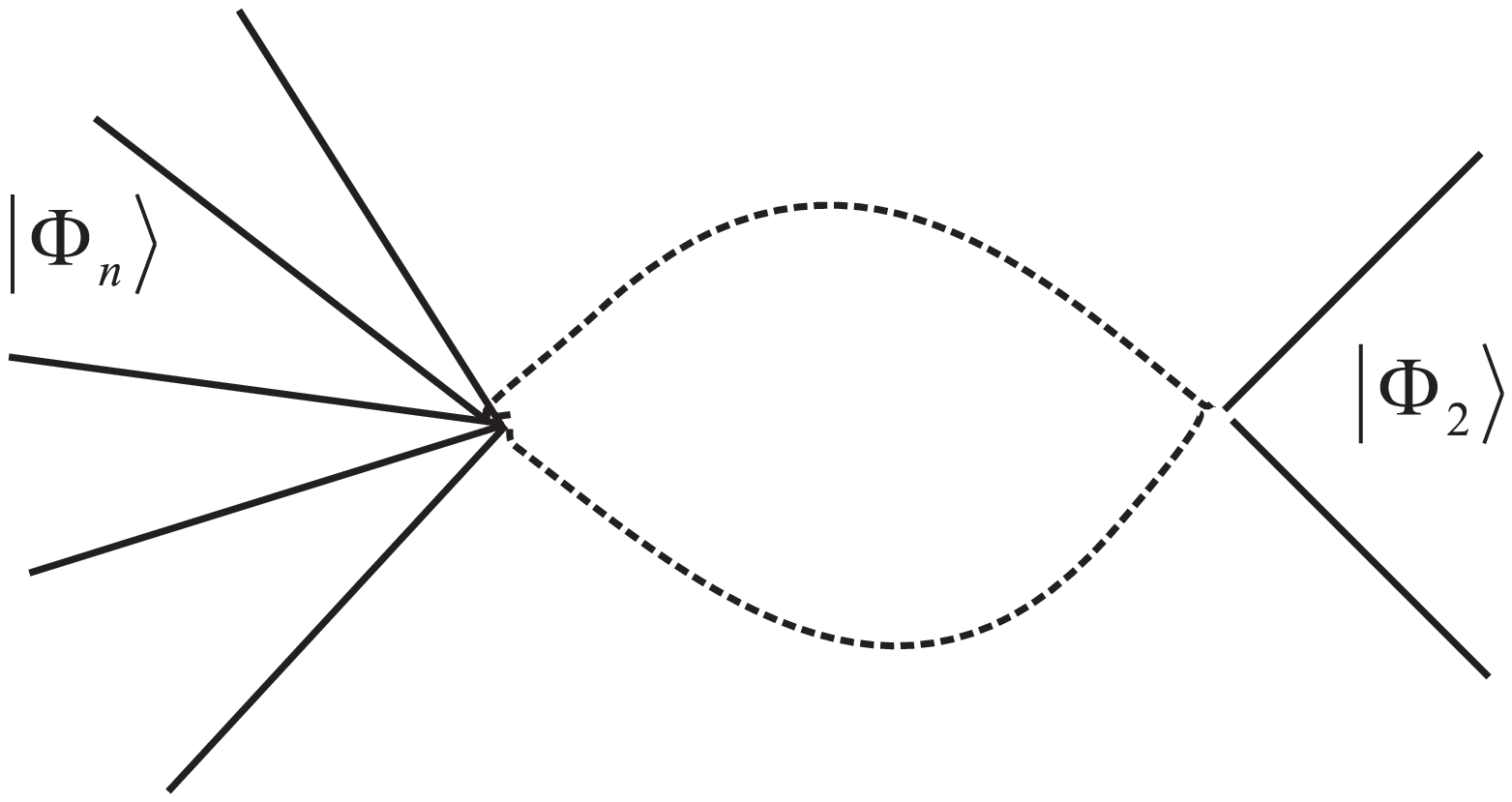}
\caption{Feynman time-symmetric twins} \label{graficof}
\end{center}
\end{figure}

Let us consider the Feynman graphs of Fig. \ref{graficof}, where
the horizontal direction ideally corresponds to the time axis, the
vertical direction represents a spatial axis, $|\Phi _{2}\rangle $
is a two-particle state and $|\Phi _{n}\rangle $ is an
$n$-particle state.

As we have argued in the previous sections, at this point there is
no substantial criterion to select the past-to-future direction on
the time axis. Thus, we cannot even consider motions (e.g.
convergence to the vertex or divergence from the vertex) along the
lines of the graph because both graphs are formally identical: one
is the temporal mirror image of the other. Moreover the
probabilities of both processes are equal:

\begin{equation}
p_{1}=|\langle \Phi _{2}|\Phi _{n}\rangle |^{2}=|\langle \Phi _{n}|\Phi
_{2}\rangle |^{2}=p_{2}  \label{33}
\end{equation}
This fact shows that the probabilities are not affected by the direction of
time. The time-symmetry of both graphs results from the time-reversal
invariance of the physical laws on which the graphs are based. If we wanted
to distinguish them, we should say that in one graph the state $|\Phi
_{2}\rangle $ is at the left (i.e. in the past) of the state $|\Phi
_{n}\rangle $, and in the second graph $|\Phi _{2}\rangle $ is at the right
(i.e. in the future) of $|\Phi _{n}\rangle $. But this argument requires a
theoretical reason to say which one of the states is at the left of the
other.

If we want to turn the merely conventional difference between the
two graphs of Fig. \ref{graficoff} into a substantial difference,
we have to consider the energy flow trough the process. In fact,
if we represent such a flow by means of arrows (they are not
fermion arrows!!!), we obtain the following figure:
\begin{figure}[h]
\begin{center}
\includegraphics[width=6cm,angle=0]{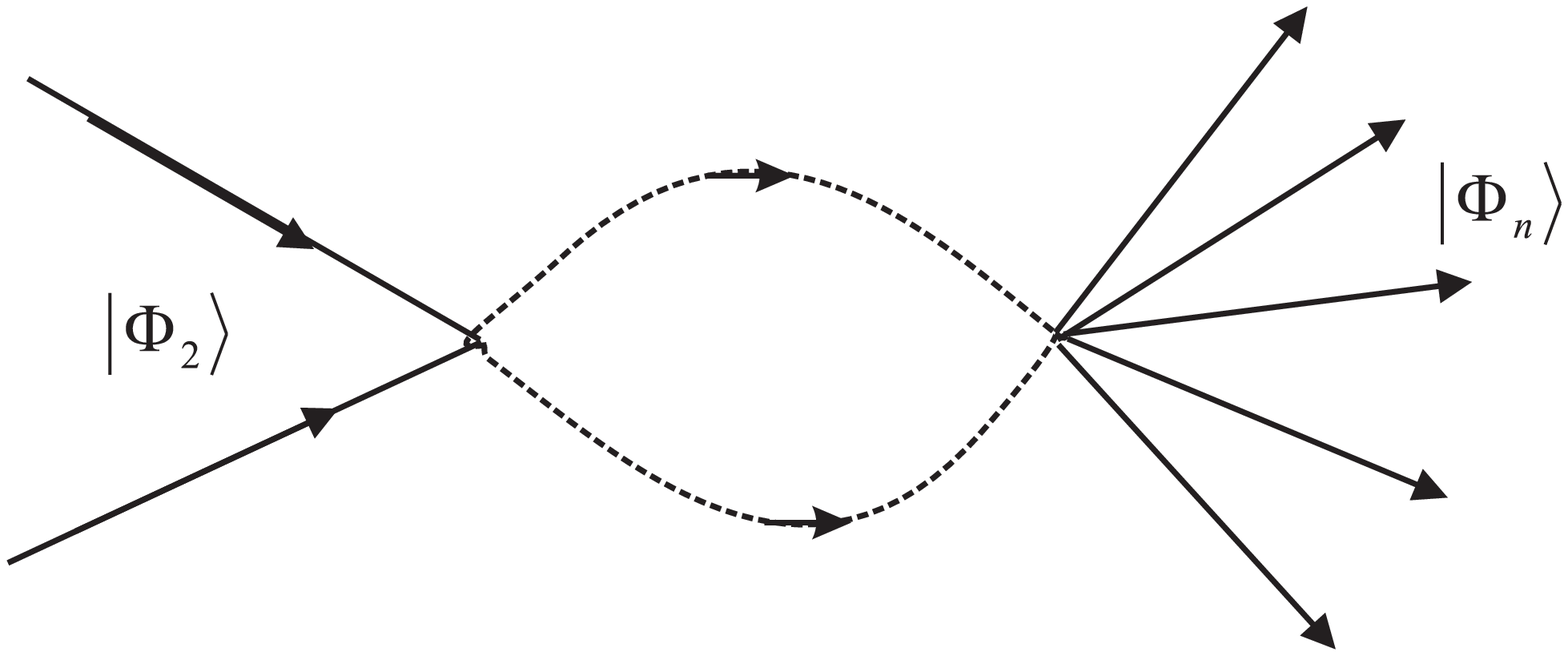} \\
\includegraphics[width=5cm,angle=0]{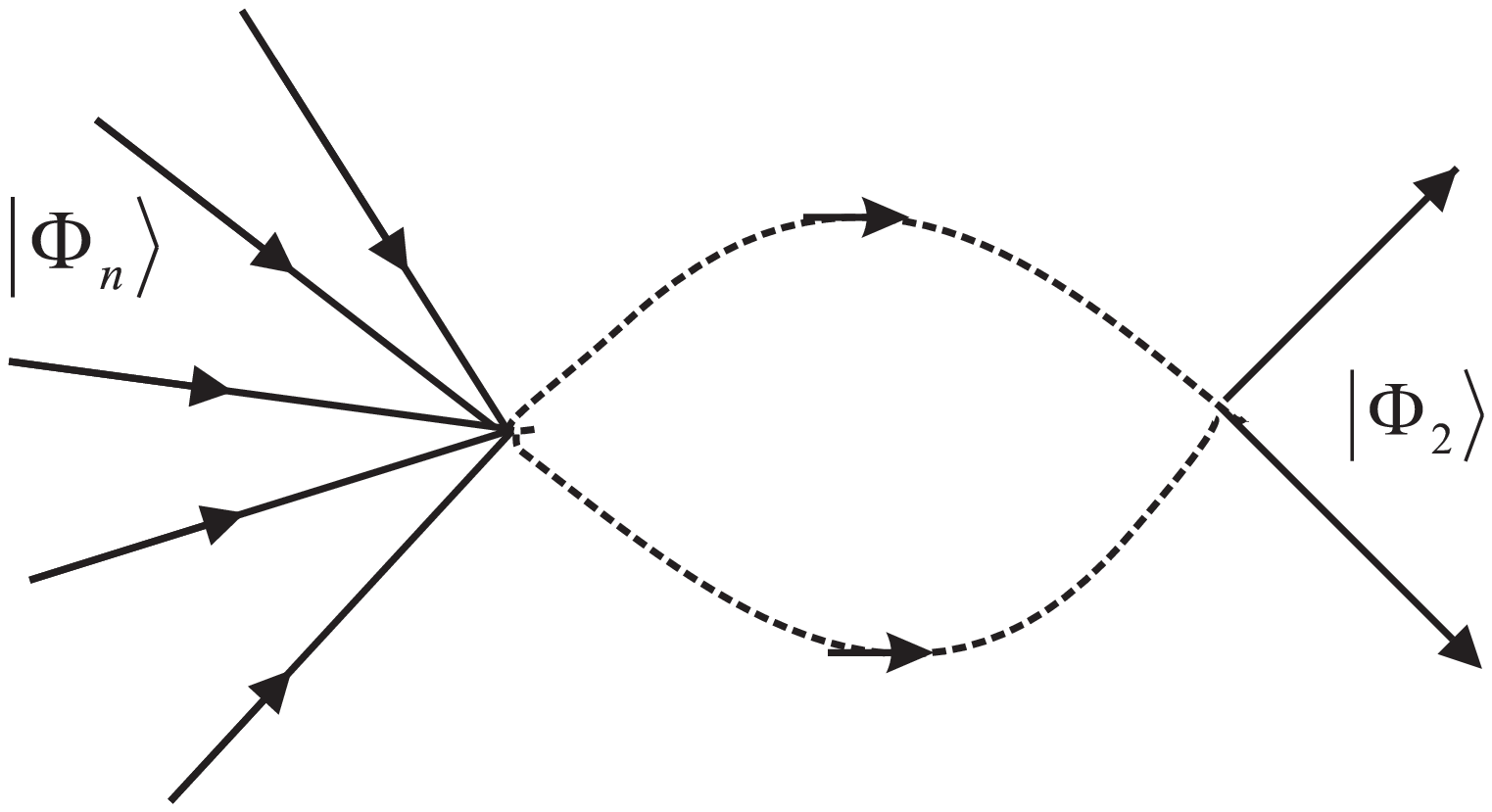}
\caption{Feynman twins with broken symmetry} \label{graficoff}
\end{center}
\end{figure}
Now both process are substantially different: in the first one two arrows
converge to the target point and $n$ arrows diverge from it, while in the
second one $n$ arrows converge and only two diverge. As we can see, the
temporal mirror image of one of the graphs is not the other: both graphs are
not formally identical because the flow of energy introduces a substantial
difference in the pair of time-symmetric twins{\it .} On the basis of this
substantial difference between the two graphs, now we can define the first
one as the typical quantum scattering process and call $|\Phi _{2}\rangle $
the ''prepared state'' and $|\Phi _{n}\rangle $ the ''detected state''. Only
on these theoretical grounds we can say that the arrow of time goes {\it %
from preparation to registration in a quantum scattering process}.\bigskip

{\bf 2. von Neumann quantum measurements\smallskip }

\begin{figure}[h]
\begin{center}
\includegraphics[width=5cm,angle=0]{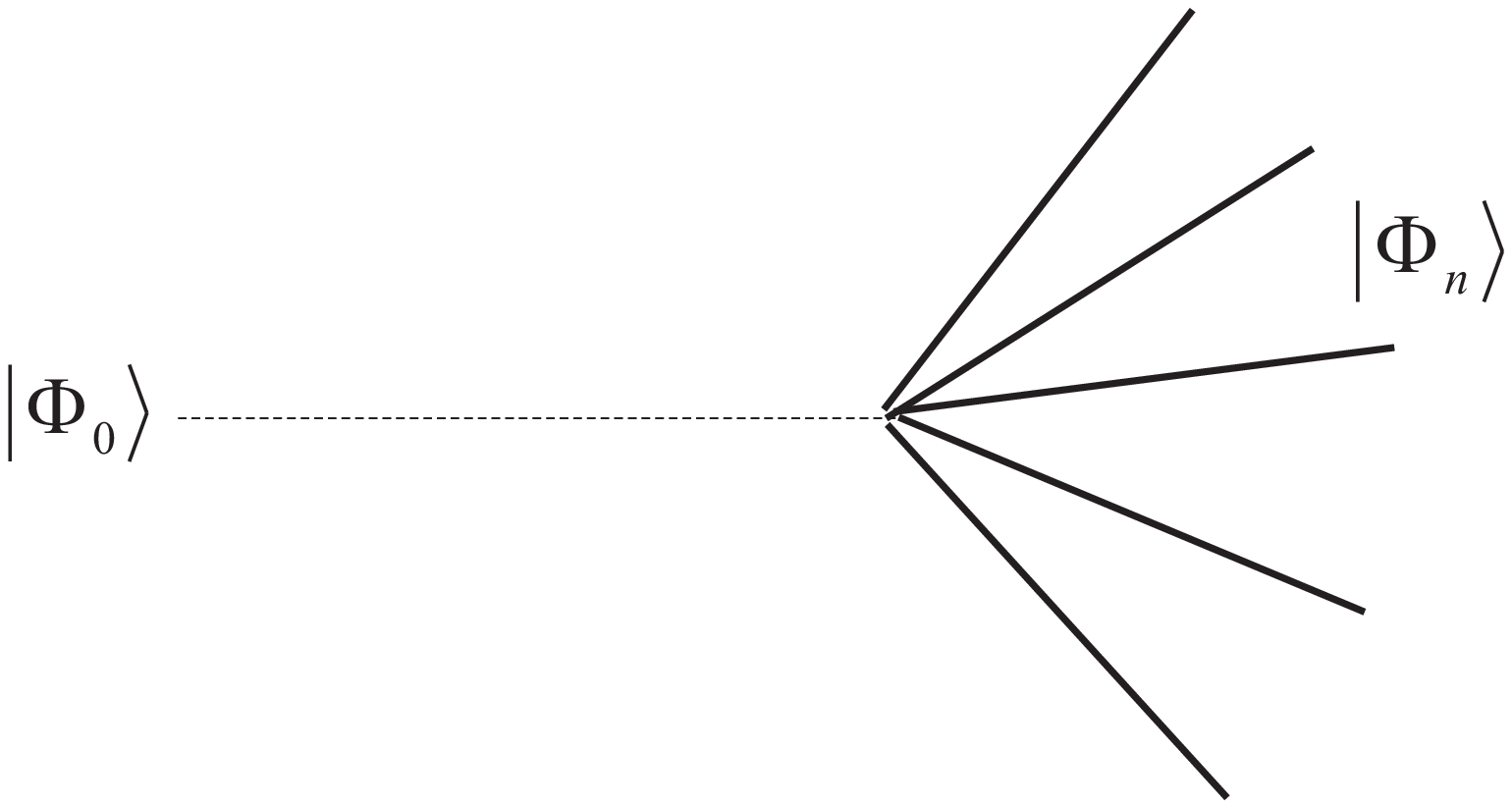} \\
\includegraphics[width=5cm,angle=0]{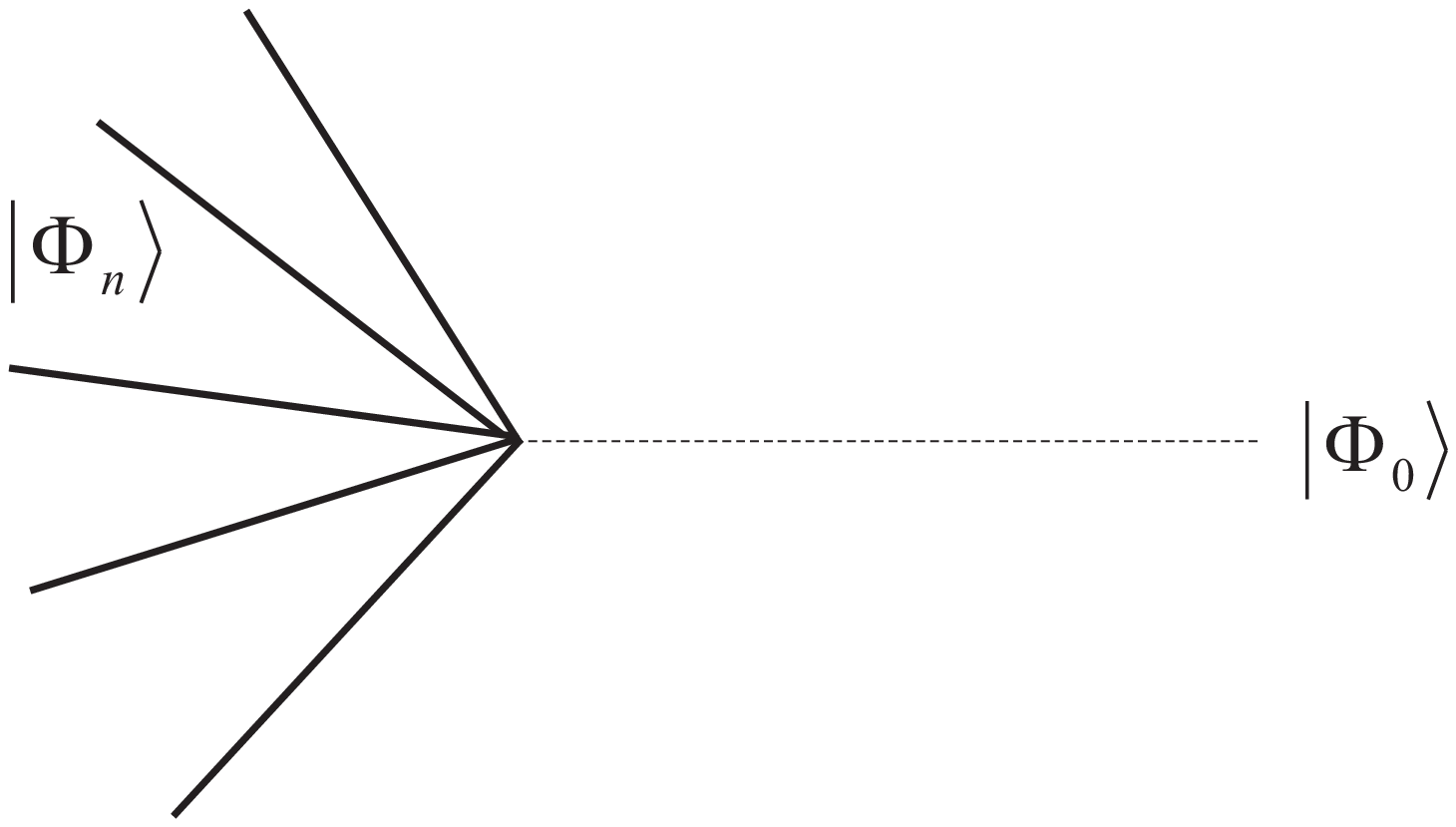}
\end{center}
\caption{Measurement time-symmetric twins} \label{graficoqm}
\end{figure}

The argument above can be easily applied to the case of quantum
measurement. Let us consider the two graphs of Fig.
\ref{graficoqm}, representing a typical von Neumann measurement,
where
\begin{equation}
|\Phi _{0}\rangle =|\varphi _{i}\rangle |A_{0}\rangle \quad \quad |\Phi
_{n}\rangle =|\varphi _{i}\rangle |A_{i}\rangle   \label{34}
\end{equation}
being $|\varphi _{i}\rangle $ the state that we want to measure, and the $%
|A_{i}\rangle $ the eigenstates of the pointer observable. In both cases,
the measurement can be performed on the basis of the correlation $|\varphi
_{i}\rangle \longleftrightarrow |A_{i}\rangle $. As in the case of the
Feynman graphs, in the measurement situation the arrow of time is usually
introduced by saying that it goes from the preparation state $|\Phi
_{0}\rangle $ to the set of measured states represented by $|\Phi
_{n}\rangle $. But, as in the previous subsection, this amounts to putting
the arrow by hand, without theoretical grounds.

\begin{figure}[h]
\begin{center}
\includegraphics[width=5cm,angle=0]{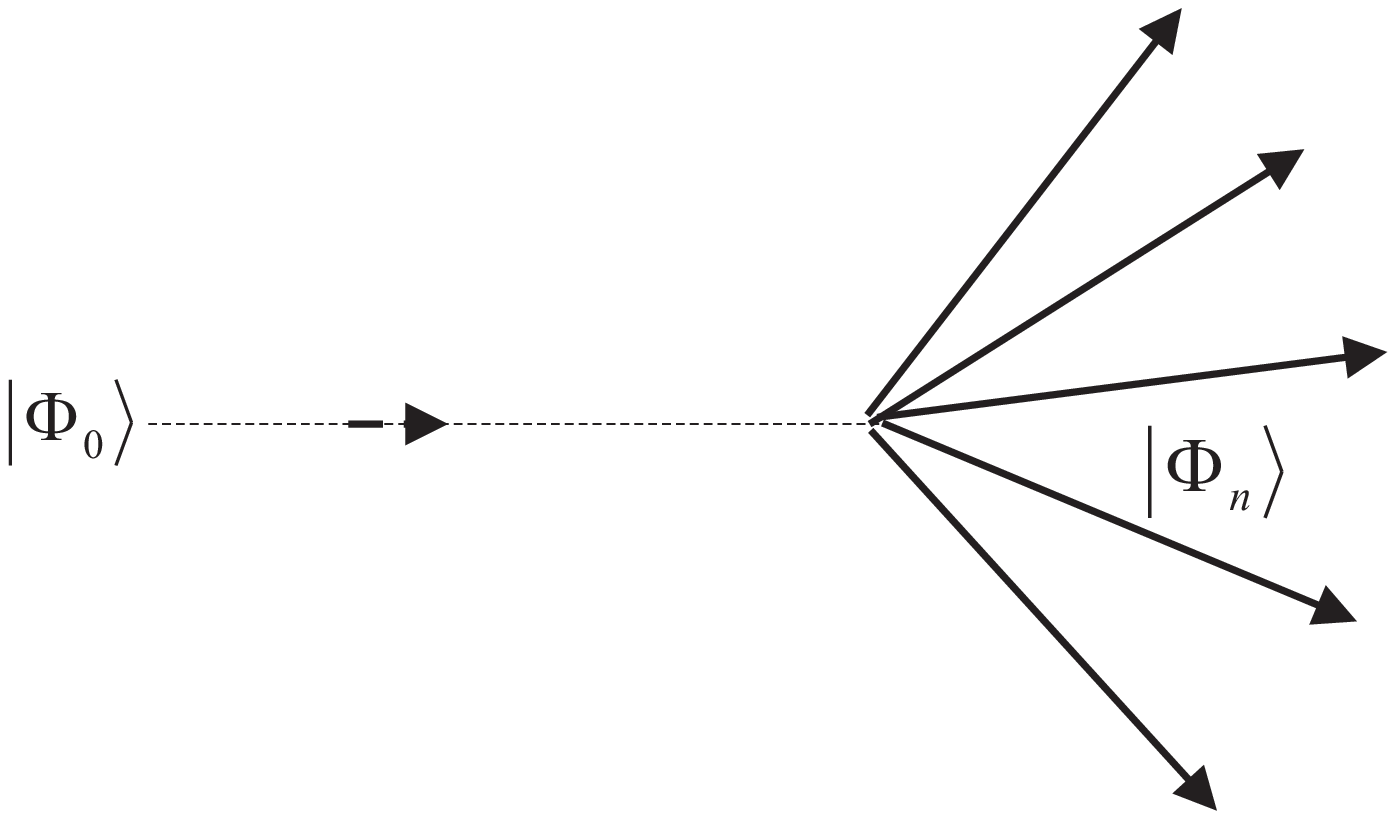} \\
\includegraphics[width=5cm,angle=0]{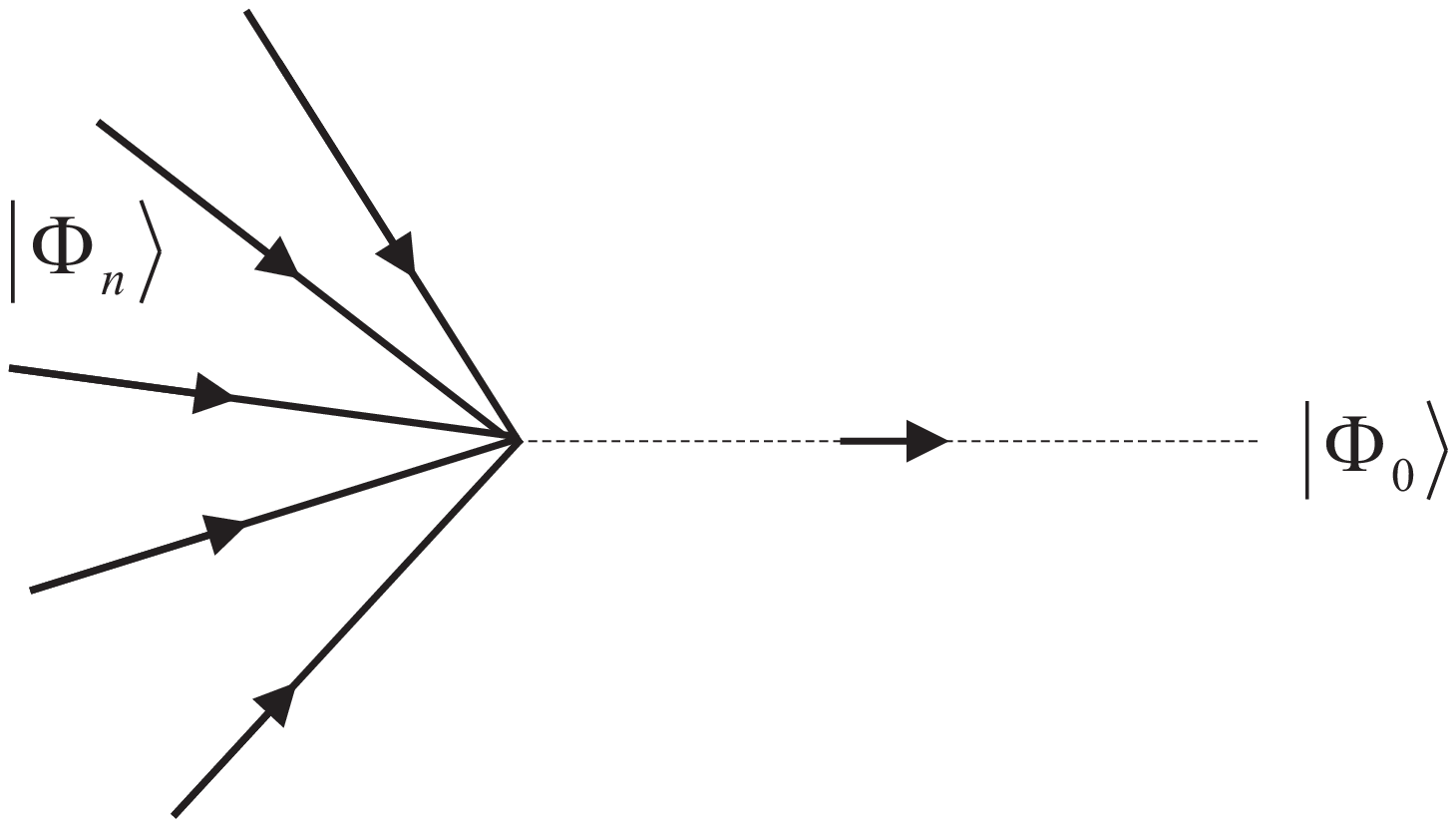}
\caption{Measurement twins with broken symmetry}
\label{graficoqmm}
\end{center}
\end{figure}

Once again, if the energy flow trough the process is considered,
the two members of the pair of graphs of Fig. \ref{graficoqm} turn
out to be substantially different and can be represented as in
Fig. \ref{graficoqmm}.
On the basis of this substantial difference, now we can define the first
graph as representing the typical quantum measurement, and call $|\Phi
_{0}\rangle $ the ''prepared state'' and $|\Phi _{n}\rangle $ the ''measured
state''. Analogously to the previous case, the arrow of time goes {\it from
preparation to registration in a quantum measurement process}.

\section{Phenomenological theories}

\subsection{Phenomenological twins}

Phenomenological theories are usually non time-reversal invariant; so, the
solution directed towards the future is taken as the physically relevant.
However, in those theories the complexity of the fundamental models is
hidden in some phenomenological coefficients that are assumed as positive;
but if these coefficients are deduced from underlying fundamental theories,
always a negative counterpart can be discovered. This means that, when a
phenomenological theory is explained in fundamental terms, the hidden
time-reversal invariance becomes manifest, and the corresponding pair of
time-symmetric twins can be identified. Let us give some examples:

\begin{itemize}
\item  {\bf The dumped oscillator}: This is{\bf \ }the paradigmatic example.
Let us consider the equation of the dumped harmonic oscillator
\begin{equation}
\ddot{x}+\omega ^{2}x+\frac{\zeta }{m}\dot{x}=0  \label{35}
\end{equation}
where $-\zeta \dot{x}$ is the viscosity term, which is opposed to the motion
if the bulk viscosity is $\zeta \geq 0$. If we make the ansatz $x=e^{\alpha
t}$, we obtain the solution
\begin{equation}
x(t)=e^{-\frac{\gamma }{2}t}e^{\pm i\widetilde{\omega }t}x(0)  \label{36}
\end{equation}
where $\gamma =\frac{\zeta }{m}\geq 0$ and $\widetilde{\omega }=\sqrt{\omega
^{2}-\left( \frac{\gamma }{2}\right) ^{2}}$ (we are just considering the
dumped oscillation case where $\omega ^{2}-\left( \frac{\gamma }{2}\right)
^{2}\geq 0$). As a consequence, the resulting evolution is a dumped motion
towards equilibrium: for $t\rightarrow \infty $, $x(t)\rightarrow 0$. The
energy obtained by this process is dissipated towards the future, e.g. in
the form of heat. This evolution is one of the phenomenological twins. The
second twin is the time-reversal version of eq. (\ref{36}), where $\zeta $
and $\gamma $ are {\it negative. }This seems strange at first sight, but the
existence of this {\it antidissipative} solution is a necessary consequence
of the time-reversal invariance of the fundamental laws underlying the
process. If in the first twin there was a flow of energy dissipated towards
the future, in the second twin the energy that amplifies the oscillations
comes {\it from the past. }

\item  {\bf The Fourier law:} The Fourier law in thermodynamics tells us
that the heat transport goes from a higher temperature region to a lower
temperature region:
\begin{equation}
{\bf J}=-K{\bf \nabla }T  \label{37}
\end{equation}

If this equation were deduced from a fundamental theory, it would result
time-reversal invariant. In turn, we know that the temperature $T$ and the
gradient ${\bf \nabla }$ do not change sign with time reversal, and that for
the flow, ${\cal T}{\bf J}=-{\bf J}$ . Therefore, if eq. (\ref{37}) has to
result time-reversal invariant, then ${\cal T}K=-K$: the negative
coefficient $K$ will lead to the second time-symmetric twin.

\item  {\bf Perturbative master equation in quantum Brownian motion: }In Paz
and Zurek$^{(40)}$, this equation is computed and the coefficient $\gamma (t)
$ is given in eq. (3.13). Since this equation is deduced from the
time-reversal invariant equations of quantum mechanics, it has to be also
time-reversal invariant. It is easy to show that, if we perform the
time-reversal $t\rightarrow -t$, we found $\gamma (-t)=-\gamma (t)$. This
means that in this case $\gamma (t)$ can be either positive or {\it negative}
and, therefore, antidissipative processes are as possible as dissipative
ones.

\item  {\bf Perturbative master equation for a two-level system coupled to a
bosonic heat bath: }In Paz and Zurek$^{(40)}$, this equation is computed and
the coefficient $f(t)$ (which plays the role of $\gamma (t)$) is given in
eq. (3.21). Again, it is easy to verify that, under the time-reversal $%
t\rightarrow -t$, we obtain $f(-t)=-f(t)$, and also in this case $f(t)$ can
be either positive or {\it negative:} antidissipative processes are as
possible as dissipative ones.

\item  {\bf Perturbative master equation for a particle coupled with a
quantum field: }In Paz and Zurek$^{(40)}$, this equation is computed (eq.
(3.25)) and the coefficient $\Gamma (x,x^{\prime },t)$ (which plays the role
$\gamma (t)$) is given in the dipole approximation of eq. (5.2). Once more,
it is easy to verify that, under the time-reversal $t\rightarrow -t$, we
obtain $\Gamma (x,x^{\prime },-t)=-\Gamma (x,x^{\prime },t)$, and also in
this case $\Gamma (x,x^{\prime },t)$ can be either positive (dissipative
processes) or {\it negative} (antidissipative processes).

\item  {\bf Quantum field theory in curved spacetime}:{\bf \ }An example
from a completely different chapter of physics comes from the theory of
fields in curved spacetime (Birrel and Davies$^{(41)}$). Let us consider a
flat FRW universe and a scalar field $\psi (t,{\bf x)=}e^{-i\frac{{\bf k}}{a}%
.{\bf x}}\varphi (t)$, where $a$ is the scale factor of the universe, ${\bf k%
}$ is the linear momentum and $\varphi (t)$ is the time evolution factor
that satisfies
\begin{equation}
\ddot{\varphi}(t)+3H\dot{\varphi}(t)+\left[ m^{2}+\left( \frac{{\bf k}}{a}%
\right) ^{2}\right] \varphi (t)=0  \label{38}
\end{equation}
where $H=\frac{\dot{a}}{a}$ is the Hubble coefficient and $m$ is the mass of
the scalar field. We see that the last equation is similar to eq. (\ref{35})
if we make the analogy $\zeta =3mH$. When $H>0$, the universe describes a
dissipative evolution in such a way that $\varphi (t)$ vanishes for $%
t\rightarrow \infty $ (Castagnino {\it et al}.$^{\text{ }(42)}$; see also
Castagnino {\it et al}.$^{(43)}$ for the similar case of fluctuations in a
FRW background). But even if $H>0$ in an expanding universe, in a {\it %
contracting} universe $H<0$. This shows that the time-reversal invariant
equations of general relativity do not exclude a negative viscosity that
leads to the second time-symmetric twin.
\end{itemize}

\subsection{Viscosity and thermal conductivity}

In this subsection we will consider how the pair of time-symmetric twins
arises in the case of the equations of a viscous fluid, when they are
deduced from the time-reversal invariant equations of classical mechanics.
When Newton's Law is applied to a small fluid volume, the Navier-Stokes
equations are obtained:

\begin{equation}
\rho \left( \frac{\partial }{\partial t}+u_{i}\frac{\partial }{\partial x_{i}%
}\right) u_{j}=F_{j}-\frac{\partial P_{ij}}{\partial x_{i}}  \label{39}
\end{equation}
where $\rho $ is the fluid density, $u_{i}$ is the velocity, $P_{ij}$ is a
potential, the $F_{j}$ are the external forces, and the $G_{j}=-\frac{%
\partial P_{ij}}{\partial x_{i}}$ are the internal forces, that is, the
forces due to the neighboring fluid elements. If we consider the case of a
non-rotational fluid, we obtain the following expression for $P_{ij}$ (see
Huang$^{(44)}$):

\begin{equation}
P_{ij}=\delta _{ij}p-\mu \left( \frac{\partial u_{i}}{\partial x_{j}}+\frac{%
\partial u_{j}}{\partial x_{i}}-\frac{2}{3}\delta _{ij}\frac{\partial u_{k}}{%
\partial x_{k}}\right)  \label{40}
\end{equation}
where $p$ is the pressure and $\mu $ is the viscosity.

Since eq. (\ref{40}) was obtained exclusively by means of classical
mechanics, it is necessarily time-reversal invariant. On this basis, we can
infer the behavior of the viscosity $\mu $ under the application of the
time-reversal operator ${\cal T}$. For the l.h.s. of eq. (\ref{40}),

\begin{equation}
{\cal T}P_{ij}=P_{ij}  \label{41}
\end{equation}
because $P_{ij}$ is a potential that verifies $G_{j}=-\frac{\partial P_{ij}}{%
\partial x_{i}}$, where the $G_{j}$ are the internal forces of the fluid. On
the other hand, for the first term of the r.h.s. of eq. (\ref{40}),
\begin{equation}
{\cal T}\delta _{ij}p=\delta _{ij}p  \label{42}
\end{equation}
since $p$ is the pressure, i.e. force per unit area. And for the second
term,
\begin{equation}
{\cal T}\left( -\mu \left( \frac{\partial u_{i}}{\partial x_{j}}+\frac{%
\partial u_{j}}{\partial x_{i}}-\frac{2}{3}\delta _{ij}\frac{\partial u_{k}}{%
\partial x_{k}}\right) \right) =-{\cal T}(\mu )\left( -\frac{\partial u_{i}}{%
\partial x_{j}}-\frac{\partial u_{j}}{\partial x_{i}}+\frac{2}{3}\delta _{ij}%
\frac{\partial u_{k}}{\partial x_{k}}\right)  \label{43}
\end{equation}
Given eqs. (\ref{41}) and (\ref{42}), this second term must be also
invariant under the application of ${\cal T}$. Therefore, the viscosity must
change as ${\cal T}\mu =-\mu $. This shows that the time-reversal invariant
fundamental laws underlying the phenomenological equations allow for either
positive and negative values of $\mu $. In the usual formulation of the
phenomenological equation, only the positive values are considered; but when
such an equation is derived from fundamental laws, the second twin that
leads to an antidissipative process becomes manifest.

We can go further to track the origin of these negative values up to the
microscopic level. Let us consider the Maxwell-Boltzmann distribution:
\begin{equation}
f=\frac{n}{(2\pi m\theta )^{\frac{3}{2}}}e^{-\frac{m}{2\theta }(v-V)^{2}}
\label{44}
\end{equation}
where $\theta =kT$ and $V$, the most probable speed of a molecule of the
gas, coincides with the maximum of $f$, $f_{\max }=v^{2}f(v)$. This
distribution applies when the gas is in equilibrium. The standard method for
studying a gas near equilibrium consists in considering a family of these
distributions in the neighborhood of any point of the gas and solving the
equations to different orders of corrections to the Maxwell-Boltzmann
function. Then, by using the transport equations for a gas with $n$
molecules per unit volume to the first order approximation of eq. (\ref{44}%
), valid for a gas close to the equilibrium state (see Huang$^{(44)}$), we
arrive to a statistical expression for the viscosity $\mu $ and for the
thermal conductivity $K$:

\begin{equation}
\mu =\frac{n\theta \lambda }{V}\qquad \qquad K=\frac{5n\theta \lambda }{2V}
\label{45}
\end{equation}
where $\lambda $ is the mean free path. But, again, if we apply
the time-reversal operator to $f_{\max }$, we obtain ${\cal
T}f_{\max }=v^{2}f(-v)$; now the most probable speed is $-V$ as we
can see in Fig. \ref{distri}:

\begin{figure}[h]
\begin{center}
\includegraphics[width=5cm,angle=0]{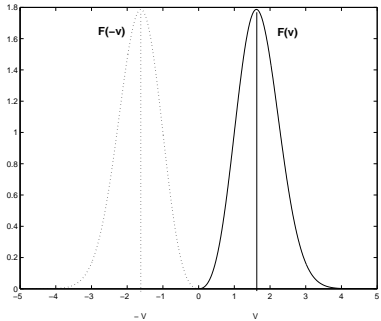}
\caption[]{Time-symmetric graphs of $f_{\max }$} \label{distri}
\end{center}
\end{figure}

Since $\theta =kT=\frac{2}{3}\epsilon $, where $\epsilon $ is the kinetic
energy of the gas particles, and
\begin{equation}
\epsilon =\frac{\int_{0}^{\infty }d^{3}p:\frac{p^{2}}{2m}f(p)}{%
\int_{0}^{\infty }d^{3}p:f(p)}  \label{46}
\end{equation}
clearly we verify eq. (\ref{3}), i.e. ${\cal T}\epsilon =\epsilon $;
therefore, ${\cal T}\theta =\theta $. Nevertheless, from eqs. (\ref{45}) we
can see that ${\cal T}K=-K$ and ${\cal T}\mu =-\mu $: both $\mu $ and $K$
change their sign under the application of the time-reversal operator,
result that again unmasks the second twin of the time-symmetric pair.

\subsection{Phenomenological thermodynamics and phenomenological entropy}

If the entropy of phenomenological thermodynamics is to be derived from
fundamental laws, the equation that makes it to grow only towards the future
(the second law) has to be one element of a pair of time-symmetric twins: as
the Eherenfests pointed out many years ago, there must exist the
time-reversed twin that makes entropy to grow towards the past. The problem
consists in discovering the second twin by appealing to the fundamental
definitions underlying the phenomenological approach.

The entropy balance equation reads
\begin{equation}
\frac{\partial \rho _{S}}{\partial t}+{\bf \nabla \cdot
J}_{S}=\sigma   \label{47}
\end{equation}
where $\rho _{S}$ is the entropy per unit volume, ${\bf J}_{S}$ is the
entropy flow and $\sigma $ is the entropy production per unit volume. If the
$X^{A}$ are the thermodynamic forces or affinities such that $X^{A}=\nabla
\gamma ^{A}$, where $\gamma ^{A}$ are the thermodynamic variables (i.e. the
coordinates of the thermodynamic space, $A,B,...=1,...,n,$ being $n$ the
number of thermodynamics variables \footnote{%
We are considering a space of just one dimension for simplicity.}), and the $%
J_{A\text{ }}$ are the thermodynamic flows, $\sigma $ reads
\begin{equation}
\sigma =\sum_{A}J_{A}X^{A}  \label{48}
\end{equation}
Then, the Onsanger-Casimir relations near equilibrium read (Castagnino {\it %
et al}.$^{\text{ }(42)}$)
\begin{equation}
J_{A}=\sum_{B}M_{AB}X^{B}  \label{49}
\end{equation}
where $M_{AB}$ is a matrix containing the constant phenomenological
coefficients (as the coefficient $\gamma $ of eq. (\ref{36}), or the bulk
viscosity $\zeta ,$ the shear viscosity $\eta ,$ the heat conduction $\chi $
and all the remaining coefficients of the previous subsection), such that
\begin{equation}
M_{AB}=L_{AB}+f_{AB},\text{ \quad }L_{AB}=L_{BA},\text{ \quad }f_{AB}=-f_{BA}
\label{50}
\end{equation}
Therefore, the entropy production results
\begin{equation}
\sigma =\sum_{AB}M_{AB}X^{A}X^{B}=\sum_{AB}L_{AB}X^{A}X^{B}  \label{51}
\end{equation}
The phenomenological second law of thermodynamic states that
\begin{equation}
\sigma \geq 0  \label{52}
\end{equation}
As a consequence, $L_{AB}$ is a positive definite matrix, that is, all the
constant phenomenological coefficients ($\gamma ,\zeta ,\eta ,\chi ,\mu ,K$)
are positive. This means that the second law describes dissipative processes
corresponding to the future-directed twin of a time-symmetric pair.

Of course, the corresponding antidissipative twin is obtained simply
changing the signs. However, the existence of such a second twin can also be
proved by considering the original definition of entropy:
\begin{equation}
dS=\frac{dH}{T}  \label{53}
\end{equation}
In this definition, ${\cal T}H=H$ (eq. (\ref{3})) and, since $T\geq 0$, $%
{\cal T}T=T$; therefore, ${\cal T}S=S$. Moreover, $S=$ $\rho _{S}V$ and,
therefore, ${\cal T}\rho _{S}=\rho _{S}$. And since ${\bf J}_{S}=\rho _{S}%
{\bf v}${\bf ,} then ${\cal T}{\bf J}_{S}=-{\bf J}_{S}$. By applying these
results to eq. (\ref{47}), we can conclude that
\begin{equation}
{\cal T}\sigma =-\sigma  \label{54}
\end{equation}
This means that, when thermodynamics is expressed in terms of fundamental
definitions, the time-symmetric twin of the second law comes to the light:
the evolutions with $\sigma <0$ are nomologically possible and, in some
cases, $L_{AB}$ is a negative definite matrix. This mirror image behavior
corresponds to the change of signs of the phenomenological coefficients $%
\gamma ,$ $\zeta ,\eta ,$ $\chi ,\mu ,K$ studied in the previous subsection.%
\footnote{%
Also in eq. (\ref{48}), ${\cal T}J_{A}=-J_{A}$ since it is a flow, and $%
{\cal T}X_{A}=X_{A}$ since it is a force or affinity; then, ${\cal T}%
M_{AB}=-M_{AB}$. And since $\gamma ,\zeta ,\eta ,\chi ,\mu $ and $K$ are
contained in $M_{AB}$, they also have to change their signs.}

If we use just the first twin, we are in the realm of phenomenological
thermodynamics; if we use both twins, we are in the realm of fundamental
thermodynamics. However, usually we only see dissipative process; thus,
something must break the time-symmetry of the twins.\bigskip

{\bf 1. Breaking the time-symmetry with the second law\smallskip }

Let an oscillator be initially in motion in a gas atmosphere at rest. The
oscillator gradually looses its energy and finally stops, while the
initially motionless molecules get in motion. This is a paradigmatic
dissipative process with factor $e^{-\gamma t}$ and $\sigma >0$. But as the
fundamental physical equations are time-reversal invariant, the opposite
process is also possible according to the laws. If initially the oscillator
were at rest but the molecules were in {\it exactly the opposite motion than
in the final state of the previous case }(i.e. their velocities were
inverted by a Maxwell demon), the evolution would be antidissipative with
factor $e^{\gamma t}$ and $\sigma <0$. It seems to be obvious that
dissipative evolutions are much more frequent that antidissipative ones,
because the latter are produced by infrequent ''conspiracies''. However,
this is not a consequence of the dynamical laws but of the initial
conditions. We might say that non-conspirative initial conditions are easy
to be produced but conspirative initial conditions (demon conditions) are
very difficult to be obtained. Nevertheless, this is just a practical
problem: we are macroscopic beings and, for this reason, moving a single
oscillator is a simple task for us, whereas endowing a great number of
molecules with the precise ''conspirative'' motion is extremely difficult
(we need the help of the microscopic Maxwell demon). This means that the
limitation is practical and not resulting from some physical law. In fact,
sometimes practical limitations can be overcome and the inversion of
velocities can be obtained, as in the case of the spin-echo experiments (see
Balliant$^{(45)}$, Levstein {\it et al}.$^{(46)}$).

Some authors base their definition of the arrow of time and their foundation
of the second law in these practical reasons, that is, the absence of
''conspiracies'' (see, for instance, Sachs$^{(12)}$). However, such a
position forces them to face a long list of well known criticisms. In fact,
since antidissipative processes are not ruled out by the fundamental laws of
physics, ''{\it A violation of irreversibility is not forbidden as a matter
of principle but because it is highly improbable'' (}Balliant$^{(45)}$,
p.408); therefore, ''{\it Irreversibility is not an absolute concept, but is
partially subjective by nature depending in the complexity of the system and
on the details and ingenuity of our observations'' (}Balliant$^{(45)}$,
p.412). This means that the appeal to practical limitations is a
non-theoretical, even a non-objective way to break the symmetry of the
phenomenological pair of time-symmetric twins.\bigskip

{\bf 2. Breaking the time-symmetry with the energy flow\smallskip }

The future-directed energy flow, expressing the global time-asymmetry of the
spacetime, supplies a theoretically grounded mean for breaking the symmetry
of the twins.

Let us begin with the case of the Fourier law where, as we have shown, the
negative coefficient $K$ leads to the second time-symmetric twin. If the
energy flow is future-directed all over the spacetime, at each point $x$ the
quadrivector ${\bf J}$, representing the heat flow, lies in the future light
semicone $C_{+}(x)$. In the case of thermodynamics, the classical limit $%
v/c\rightarrow 0$ has to be applied; as a consequence, the semicone $%
C_{+}(x) $ becomes a future semiplane $P_{+}(x)$, but this limit does not
affect the orientation of ${\bf J}$. Then, although the time-reversal
invariant fundamental laws lead to a pair of time-symmetric twins
corresponding to the two possible orientations of ${\bf J}$, the energy flow
breaks the symmetry by selecting the quadrivector ${\bf J}$ corresponding to
$P_{+}(x)$; therefore, it explains the positive value of the coefficient $K$
and, {\it a fortiori}, the second law of thermodynamics.

From a more general viewpoint, we have seen that the time-symmetric twin of
the second law $\sigma \geq 0$ is $\sigma \leq 0$: when $\sigma >0$, the
process produces an energy flow directed towards the future; when $\sigma <0$%
, the process requires an energy flow pumped from the past. Let us note
that, if we did not have a criterion for defining the past-to-future
direction of the energy flow, the previous assertions would be senseless:
from an ''atemporal'' viewpoint that does not commits a {\it petitio
principii} by presupposing a privileged direction of time, ''past'' and
''future'' are only conventional labels (see Fig. 6). On the contrary, with
the energy flow pointing to the same direction all over the spacetime, we
can legitimately say that $\sigma >0$ corresponds to a dissipative decaying
process evolving from non-equilibrium to equilibrium as $e^{-\gamma t}$, and
$\sigma <0$ corresponds to an antidissipative growing process evolving from
equilibrium to non-equilibrium as $e^{\gamma t}$. The two processes, which
in principle are only conventionally different, turn out to be substantially
different due to the future-directed energy flow that locally expresses the
global time-asymmetry of the universe (see Fig. 7).\bigskip

\begin{figure}[h]
\begin{center}
\includegraphics[width=5cm,angle=0]{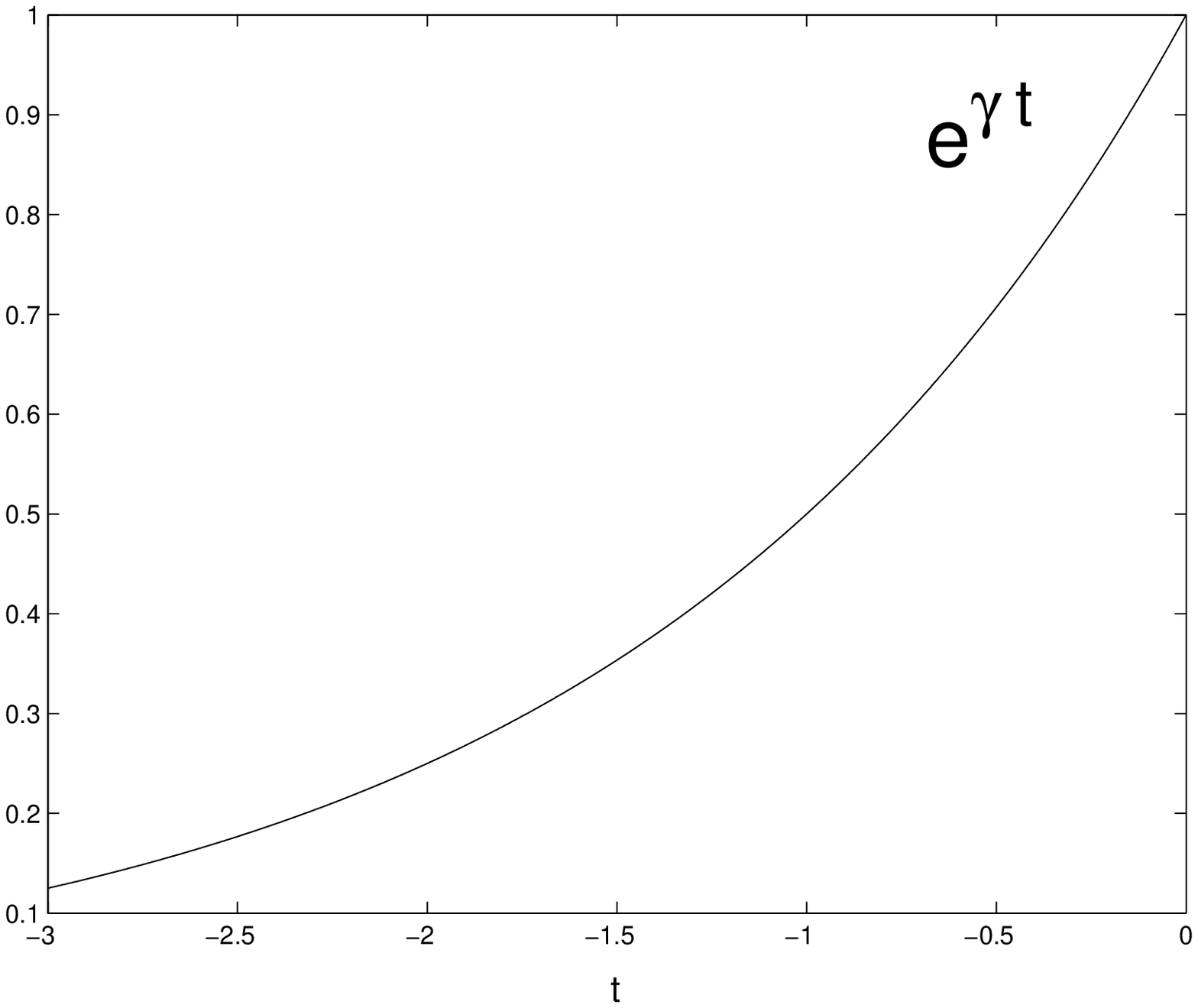}
\includegraphics[width=5cm,angle=0]{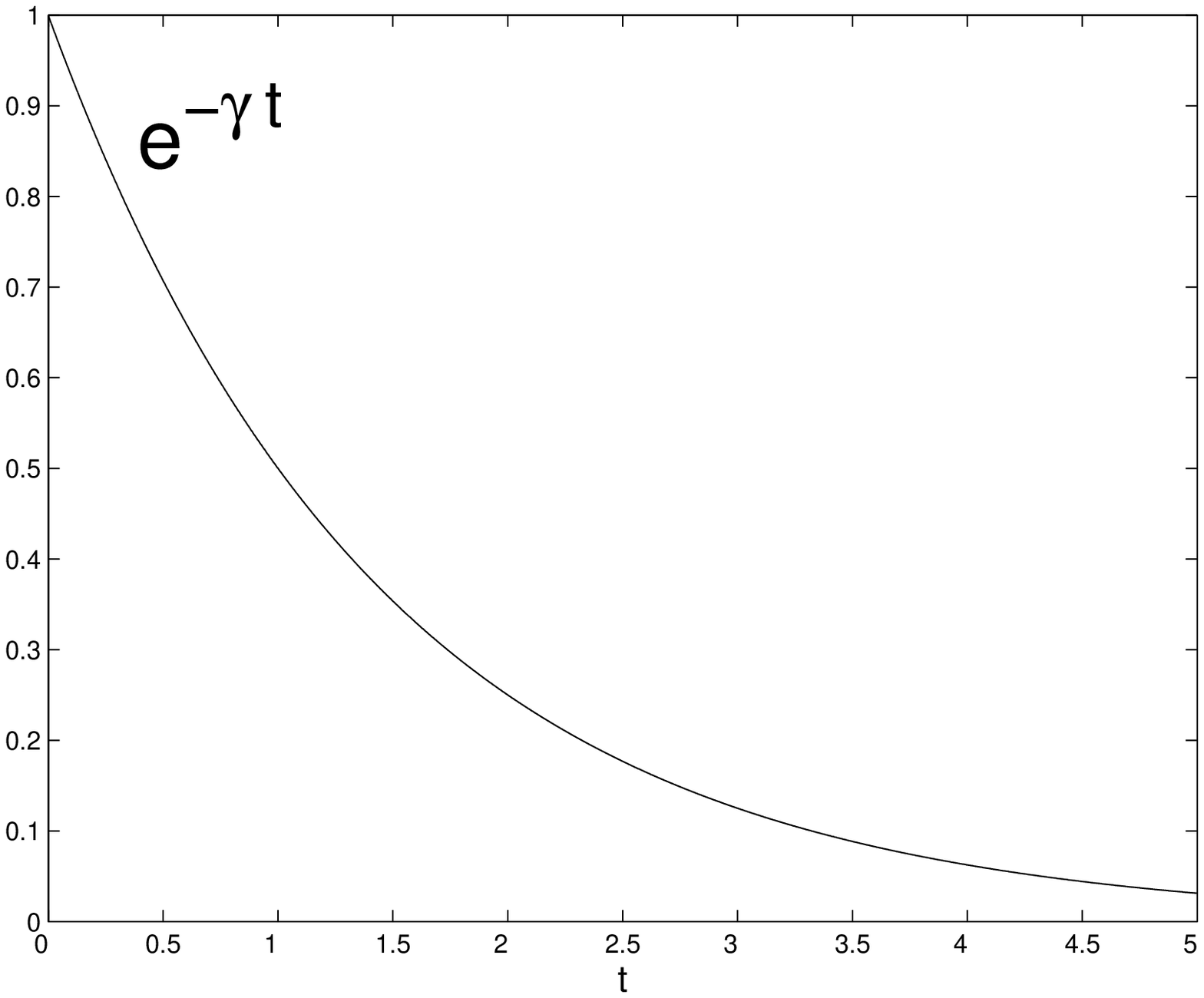}
\caption[]{Time-symmetric twins} \label{figura6}
\end{center}
\end{figure}
\begin{figure}[h]
\begin{center}
\includegraphics[width=5cm,angle=0]{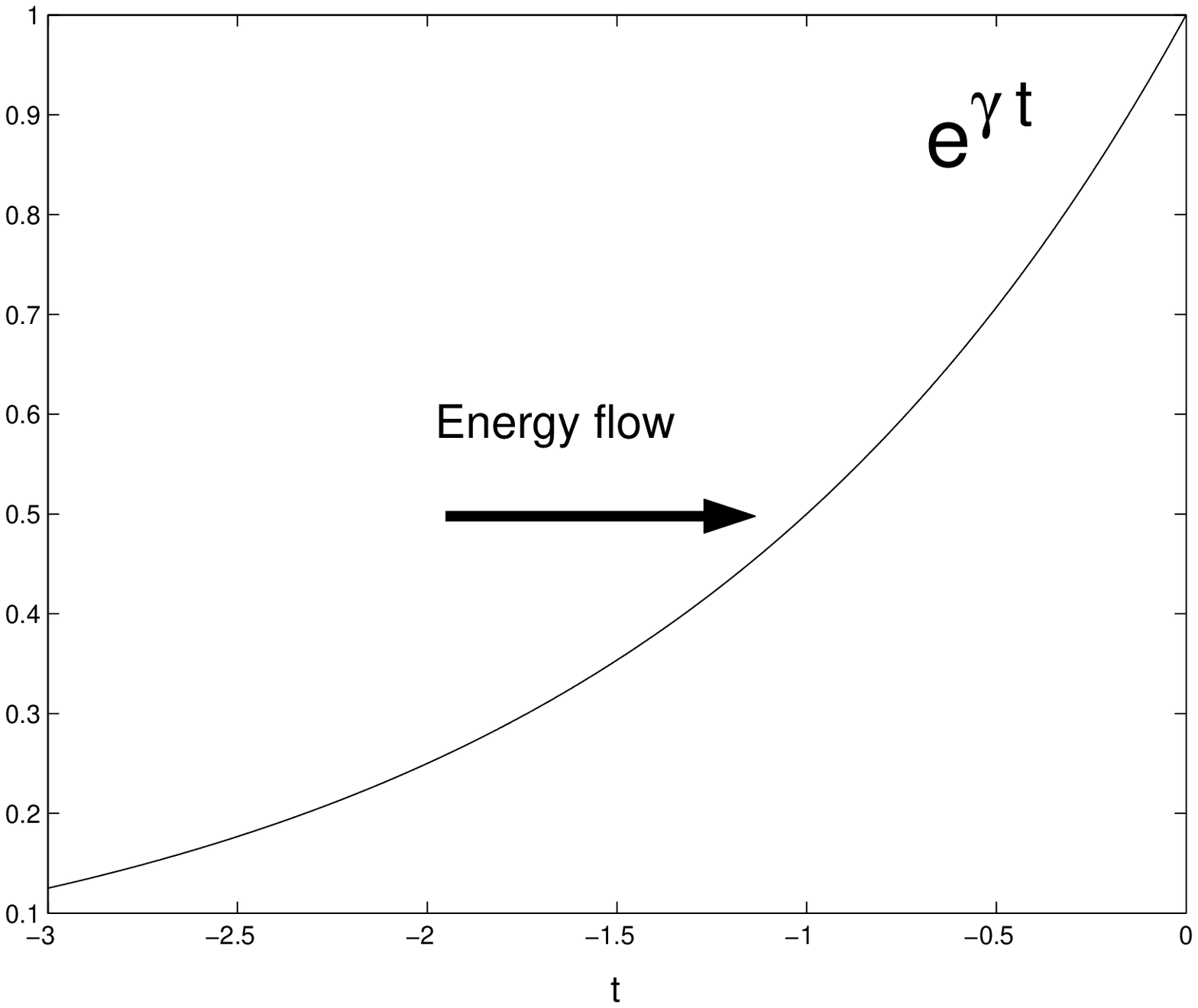}
\includegraphics[width=5cm,angle=0]{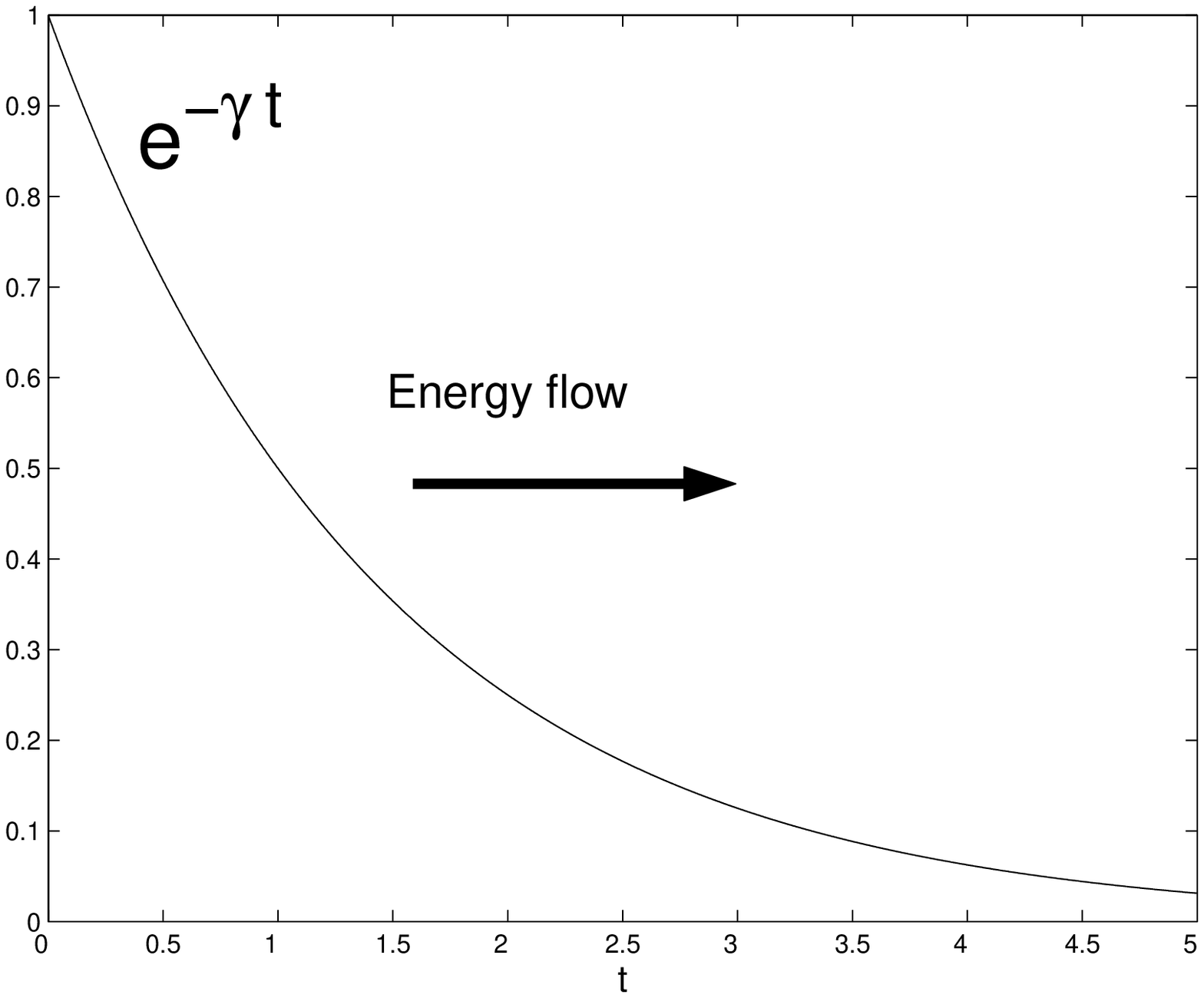}
\caption[]{ Time-symmetric twins with broken symmetry}
\label{figura7}
\end{center}
\end{figure}

\bigskip

\subsection{Entropy and energy-momentum tensor: relativistic imperfect fluids
}

Finally, let us consider how the energy-momentum tensor and the entropy are
related in the case of relativistic imperfect fluids, as a further example
of time-symmetric twins whose symmetry can be broken by a future-directed
energy flow. For a universe containing a relativistic imperfect fluid, the
energy-momentum tensor reads
\begin{equation}
T^{\mu \nu }=pg^{\mu \nu }+(p+\rho )U^{\mu }U^{\nu }+\Delta T^{\mu \nu }
\label{55}
\end{equation}
where $p$ is the pressure, $g^{\mu \nu }$ is the metric tensor, $\rho $ is
the energy-matter density, $U^{\mu }$ is the absolute velocity of the fluid,
and $\Delta T^{\mu \nu }$ is a term due to the imperfection of the fluid. In
a comoving (free falling) frame, $U^{0}=1$, $U^{i}=0$, $\Delta T^{00}=0$,
and the corresponding eq. (\ref{48}) reads
\begin{equation}
\sigma =-\left( \frac{1}{T}\dot{U}_{i}+\frac{1}{T^{2}}\frac{\partial T}{%
\partial x^{i}}\right) \Delta T^{i0}-\frac{1}{T}\frac{\partial U_{i}}{%
\partial x^{j}}\Delta T^{ij}  \label{56}
\end{equation}
where $T$ is the absolute temperature, $i,j...=1,2,3$ are the spatial
indices, and the Einstein summation convention is used as before. From this
equation we want to obtain $\Delta T^{i0}$ and $\Delta T^{ij}$. But here we
will not appeal to the traditional argument, which relies on the second law $%
\sigma \geq 0$ (see Weinberg$^{(47)}$). On the contrary, we will use a
fundamental law as eq. (\ref{48}), which will bring to the light the second
element of the pair of time-symmetric twins; the second law will be a
consequence of the symmetry breaking in the pair.

Let us define
\begin{equation}
\Delta T_{0}^{ij}=\delta ^{ij}\Delta T_{k}^{k},\qquad \Delta
T_{1}^{ij}=\Delta T^{ij}-\Delta T_{0}^{ij}  \label{57}
\end{equation}
$\Delta T_{0}^{ij}$ and $\Delta T_{1}^{ij}$ are the two irreducible
components of the symmetric tensor $\Delta T^{ij}$ under the spatial
rotation group $SO(3)$, while $\Delta T^{i0}$ is an irreducible vector under
the same group. The thermodynamic space is flat or, at least, locally flat
(near equilibrium). Therefore, since the theory must be invariant under the
rotation of $SO(3)$, the matrix $M_{AB}=M_{(\mu \nu )(\kappa \lambda )}$
must be spherically symmetric for each irreducible component, and eq. (\ref
{51}) reads
\begin{equation}
\sigma =M_{(\mu \nu )(\kappa \lambda )}\Delta T^{\mu \nu }\Delta T^{\kappa
\lambda }=\chi \Delta T^{i0}\Delta T_{i0}+\mu \Delta T_{0}^{ij}\Delta
T_{ij0}+\eta \Delta T_{1}^{ij}\Delta T_{ij1}  \label{58}
\end{equation}
where we have attributed an arbitrary scalar $\chi $, $\mu $, and $\eta $ to
each component. This equation has to be satisfied by any arbitrary values of
$\chi $, $\mu $, and $\eta $; then, by means of eq. (\ref{56}) we obtain
\begin{equation}
\Delta T^{i0}=-\chi \left( T\dot{U}_{i}+\frac{\partial T}{\partial x^{i}}%
\right) \qquad \qquad \Delta T^{ij}=-\eta \left( \frac{\partial U_{i}}{%
\partial x^{j}}+\frac{\partial U_{j}}{\partial x^{i}}-\frac{2}{3}\overline{%
\nabla }.\overline{U}\delta _{ij}\right) -\mu \overline{\nabla }.\overline{U}%
\delta _{ij}  \label{59}
\end{equation}

Having obtained $\Delta T^{i0}$ and $\Delta T^{ij}$ without appealing to the
second law $\sigma \geq 0$, now we can deduce the expression for $\sigma $
from eqs. (\ref{56}) and (\ref{59}):

\begin{equation}
\sigma =\frac{\chi }{T^{2}}\left( \overline{\nabla }T+T\dot{\overline{U}}%
\right) ^{2}+\frac{\eta }{2T}\left( \frac{\partial U_{i}}{\partial x^{j}}+%
\frac{\partial U_{j}}{\partial x^{i}}-\frac{2}{3}\overline{\nabla }.%
\overline{U}\delta _{ij}\right) \left( \frac{\partial U_{i}}{\partial x^{j}}+%
\frac{\partial U_{j}}{\partial x^{i}}-\frac{2}{3}\overline{\nabla }.%
\overline{U}\delta _{ij}\right) +\frac{\mu }{T}\left( \overline{\nabla }.%
\overline{U}\right) ^{2}  \label{60}
\end{equation}
Up to this point, we have made no assumptions about the values of the
coefficients $\chi $, $\mu $ and $\eta $; then, they can be either positive,
leading to $\sigma >0$, or negative, leading to $\sigma <0$: both
situations, one the temporal mirror image of the other, are nomologically
possible according to the fundamental laws.

Also in this case, the symmetry can be broken by means of the energy flow.
Once we have established a substantial difference between the two directions
of time and we have used, following the traditional convention, the label
''future'' for the direction of the energy flow, we can say that:

\begin{itemize}
\item  For {\it dissipative} processes, $\chi >0$, $\mu >0$ and $\eta >0$;
as a consequence, since the ''geometrical'' factors between parenthesis in
eq. (\ref{60}) are all non negative, then $\sigma >0$ (the second law).

\item  For {\it antidissipative} processes, $\chi <0$, $\mu <0$ and $\eta <0$%
; as a consequence, $\sigma <0$. In the regions of the universe where this
condition holds, the second law will be not locally valid.
\end{itemize}

\subsection{The status of the second law}

We can summarize the results of this section by saying that, according to
the fundamental laws of physics, either dissipative processes with $\sigma
>0 $ and antidissipative processes with $\sigma <0$ are nomologically
possible. The symmetry between both in principle formally identical
situations is broken only by an energy flow which points to the same
direction all over the universe and expresses the global time-asymmetry of
the spacetime.

It seems quite clear that this way of breaking the symmetry in the pair of
time-symmetric twins is theoretically grounded and not relying on merely
contingent practical limitations. It is also completely general, since it
can be applied to any pair of time-symmetric twins of physics. Furthermore,
since the energy flow points to the same direction all over the spacetime,
this symmetry breaking accounts for the otherwise unexplained fact that the
different arrows of time, defined in the different chapters of physics
(electromagnetic arrow, thermodynamic arrow, cosmological arrow, etc.), all
point to the same time direction.

This conclusion allows us to assess the status of the second law of
thermodynamics. As we have seen, when arguments are based exclusively on
fundamental laws, pairs of time-symmetric twins appear in all the chapters
of physics. In the particular case of phenomenological thermodynamics, the
twin of the second law can also be discovered. So, the traditional second
law $\sigma \geq 0$ only arises when the time-symmetry is broken by the
future-directed energy flow. But this way of breaking the symmetry is common
to all the pairs of time-symmetric twins, from electromagnetism to quantum
field theory. Therefore, the second law is not endowed with a privileged
character with respect to the arrow of time, as usually supposed: the
thermodynamic arrow, as all the other arrows, is a consequence of the global
time-asymmetry of the universe. In this sense, {\it the second law can be
inferred on the basis of global considerations}, in the same way as the
irreversible evolutions of quantum mechanics or the non time-reversal
invariant postulate of quantum field theory.

\section{Conclusions}

In this paper we have completed the following tasks:

\begin{itemize}
\item  We have disentangled the concepts of time-reversal invariance,
irreversibility and arrow of time, dissipating the usual confusions between
the problem of irreversibility and the problem of the arrow of time.

\item  We have defined the arrow of time as the global time-asymmetry of
spacetime, which is locally expressed as a future-directed energy flow all
over the universe.

\item  We have shown how, in different chapters of physics, the
time-reversal invariant fundamental laws lead to pairs of time-symmetric
twins whose elements are only conventionally different in the light of such
laws.

\item  We have shown how the future-directed energy flow is what breaks the
symmetry of all the pairs of time-symmetric twins of physics, giving rise to
the different arrows of time traditionally treated in the literature on the
subject.
\end{itemize}

With this work we have tried to contribute to the resolution of the problem
of the direction of time, one of the most longstanding debates on the
conceptual foundations of theoretical physics.

\section{Acknowledgments}

We are particularly grateful to Huw Price, who urged us to complete our work
by considering entropy. This work was partially supported by grants of the
Buenos Aires University, the CONICET and the FONCYT of Argentina.

\bigskip

{\center{\bf References}}

\bigskip

\begin{enumerate}
\item  M. Castagnino, O. Lombardi and L. Lara, ''The global arrow of time as
a geometrical property of the universe'', {\it Foundations of Physics} {\bf %
33}, 877-912 (2003).

\item  M. Castagnino, L. Lara and O. Lombardi, ''The cosmological origin of
time-asymmetry'', {\it Classical and Quantum Gravity }{\bf 20}, 369-391
(2003).

\item  M. Castagnino, L. Lara and O. Lombardi, ''The direction of time: from
the global arrow to the local arrow'', {\it International Journal of
Theoretical Physics} {\bf 42}, 2487-2504 (2003).

\item  M. Castagnino and O. Lombardi, ''The generic nature of the global and
non-entropic arrow of time and the double role of the energy-momentum
tensor'', {\it Journal of Physics A (Mathematical and General)} {\bf 37},
4445-4463 (2004).

\item  M. Castagnino and O. Lombardi, ''A global and non-entropic approach
to the problem of the arrow of time'', in A. Reimer (ed.), {\it Spacetime
Physics Research Trends. Horizons in World Physics} (Nova Science, New York,
2005).

\item  M. Castagnino and O. Lombardi, ''The global non-entropic arrow of
time: from global geometrical asymmetry to local energy flow'', {\it Synthese%
}, forthcoming.

\item  D. Albert, {\it Time and Chance} (Harvard University Press, Cambridge
MA, 2001).

\item  J. Earman, ''What time reversal invariance is and why it matters'',
{\it International Studies in the Philosophy of Science} {\bf 16}, 245-264
(2002).

\item  P. D. Lax and R. S. Phillips,{\it \ Scattering Theory (}Academic
Press, New York, 1979).

\item  M. Tabor, {\it Chaos and Integrability in Nonlinear Dynamics} (John
Wiley and Sons, New York, 1989).

\item  R. Penrose, ''Singularities and time asymmetry'', in S. Hawking and
W. Israel (eds.), {\it General Relativity, an Einstein Centenary Survey }%
(Cambridge University Press, Cambridge, 1979).

\item  R. G. Sachs, {\it The Physics of Time-Reversal} (University of
Chicago Press, Chicago, 1987).

\item  H. Price, {\it Time's Arrow and Archimedes' Point} (Oxford University
Press, Oxford, 1996).

\item  P. Ehrenfest and T. Ehrenfest, {\it The Conceptual Foundations of the
Statistical Approach in Mechanics} (Cornell University Press, Ithaca, 1959,
original 1912).

\item  S. Brush, {\it The Kind of Motion We Call Heat} (North Holland,
Amsterdam, 1976).

\item  L. Boltzmann, {\it Annalen der Physik} {\bf 60, }392-398 (1897).

\item  H. Reichenbach, {\it The Direction of Time} (University of California
Press, Berkeley, 1956).

\item  R. P. Feynman, R. B. Leighton and M. Sands, {\it The Feynman Lectures
on Physics, Vol. 1 (}Addison-Wesley{\it , }New York, 1964).

\item  P. C. Davies, {\it The Physics of Time Asymmetry }(University of
California Press, Berkeley, 1974).

\item  P. C. Davies, ''Stirring up trouble'', in J. J. Halliwell, J.
Perez-Mercader and W. H. Zurek (eds.), {\it Physical Origins of Time
Asymmetry} (Cambridge University Press, Cambridge, 1994).

\item  J. Earman, {\it Philosophy of Science} {\bf 41, }15-47 (1974).

\item  S. Hawking and J. Ellis, {\it The Large Scale Structure of Space-Time
}(Cambridge University Press, Cambridge, 1973).

\item  B. F. Schutz, {\it Geometrical Methods of Mathematical Physics}
(Cambridge University Press, Cambridge, 1980).

\item  A. Gr\"{u}nbaum, {\it Philosophical Problems of Space and Time}
(Reidel, Dordrecht, 1973).

\item  L. Sklar, {\it Space, Time and Spacetime} (University of California
Press, Berkeley, 1974).

\item  O. Penrose and I. C. Percival, ''The direction of time'', {\it %
Proceedings of the Physical Society} {\bf 79,} 605-616 (1962).

\item  R. Caldwell, M. Kamionkowski and N. Weinberg, ''Phantom energy and
cosmic doomsday'', {\it Physical Review Letters} {\bf 91}, 071301 (2003).

\item  A. Bohm and M. Gadella, {\it Dirac Kets, Gamow Vectors, and Gel'fand
Triplets (}Springer-Verlag, Berlin, 1989).

\item  A. Bohm, I. Antoniou and P. Kielanowski, ''The
preparation/registration arrow of time in quantum mechanics'', {\it Physics
Letters A} {\bf 189}, 442-448 (1994).

\item  A. Bohm, I. Antoniou and P. Kielanowski, ''A quantum mechanical arrow
of time and the semigroup time evolution of Gamow vectors'',{\it \ Journal
of Mathematical Physics} {\bf 36}, 2593-2604 (1994).

\item  A. Bohm, S. Maxson, M. Loewe, and M. Gadella, ''Quantum mechanical
irreversibility'', {\it Physica A} {\bf 236}, 485-549 (1997).

\item  A. Bohm and S. Wickramasekara, ''The time reversal operator for
semigroup evolutions'', {\it Foundations of Physics} {\bf 27}, 969-993
(1997).

\item  M. Castagnino, M. Gadella and O. Lombardi, ''Time's arrow and
irreversibility in time-asymmetric quantum mechanics'', {\it International
Studies in the Philosophy of Science} {\bf 19}, 223-243 (2005).

\item  M. Castagnino, M. Gadella and O. Lombardi, ''Time-reversal,
irreversibility and arrow of time in quantum mechanics'', {\it Foundations
of Physics} {\bf 36}, 407-426 (2006).

\item  N. Bogoliubov, A. A. Logunov and I. T. Todorov, {\it Axiomatic
Quantum Field Theory} (Benjamin-Cummings, Reading MA, 1975).

\item  P. Roman, {\it Introduction to Quantum Field Theory} (Wiley, New
York, 1969).

\item  R. Haag, {\it Local Quantum Physics. Fields, Particles, Algebras}
(Springer, Berlin, 1996).

\item  S. Weinberg, {\it The Quantum Theory of Fields} (Cambridge University
Press, Cambridge, 1995).

\item  M. Visser, {\it Lorentzian Wormholes} (Springer-Verlag, Berlin, 1996).

\item  J. P. Paz and W. H. Zurek, ''Environment-induced decoherence and the
transition from quantum to classical'', in Dieter Heiss (ed.), {\it Lecture
Notes in Physics, Vol. 587} (Springer, Heidelberg-Berlin, 2002).

\item  N. Birrell and P. Davies, {\it Quantum Fields in Curved Space}
(Cambridge University Press, Cambridge, 1982).

\item  M. Castagnino, H. Giacomini and L. Lara, ''Dynamical properties of
the conformally coupled flat FRW model'', {\it Physical Review D} {\bf 61},
107302 (2000).

\item  M. Castagnino, J. Chavarriga, L. Lara and M. Grau, ''Exact solutions
for the fluctuations in a flat FRW universe coupled to a scalar field'',
{\it International Journal of Theoretical Physics} {\bf 41}, 2027-2035
(2002).

\item  K. Huang, {\it Statistical Mechanics} (John Wiley and Sons, New York,
1987).

\item  R. Balliant, {\it From Microphysics to Macrophysics: Methods and
Applications of Statistical Physics} (Springer, Heidelberg, 1992).

\item  P. R. Levstein, G. Usaj and H. M. Pastawski, ''Attenuation of
polarization echoes in nuclear magnetic resonance: A study of the emergence
of dynamical irreversibility in many-body quantum systems'', {\it Journal of
Chemical Physics} {\bf 108}, 2718-2724 (1998).

\item  S. Weinberg, {\it Gravitation and Cosmology} (John Wiley and Sons,
New York, 1972).
\end{enumerate}

\end{document}